\documentclass[iop]{emulateapj}
\usepackage{amsmath}

\usepackage{color}

\shorttitle{Mildly-Hierarchical Triples}
\shortauthors{Bhaskar et al.}

\begin{document}

\title{Mildly-Hierarchical Triple Dynamics and Applications to the Outer Solar System}

\author{Hareesh Bhaskar \altaffilmark{1}, Gongjie Li \altaffilmark{1}, Sam Hadden \altaffilmark{2}, Matthew J. Payne \altaffilmark{2}, Matthew J. Holman \altaffilmark{2}}
\affil{$^1$ Center for Relativistic Astrophysics, School of Physics, Georgia Institute of Technology, Atlanta, GA 30332, USA}
\affil{$^2$ Harvard-Smithsonian Center for Astrophysics, The Institute for Theory and Computation, 60 Garden Street, Cambridge, MA 02138, USA}



\begin{abstract}
Three-body interactions are ubiquitous in astrophysics. For instance, Kozai-Lidov oscillations in hierarchical triple systems have been studied extensively and applied to a wide range of astrophysical systems. However, mildly-hierarchical triples also play an important role, but they are less explored. In this work we consider the secular dynamics of a test particle in a mildly-hierarchical configuration. We find the limit within which the secular approximation is reliable when the outer perturber is in a circular orbit. In addition, we present resonances and chaotic regions using surface of sections, and characterize regions of phase space that allow large eccentricity and inclination variations. Finally, we apply the secular results to the outer solar system. We focus on the distribution of extreme trans-neptunian objects (eTNOs) under the perturbation of a possible outer planet (Planet-9), and find that in addition to a low inclination Planet-9, a polar or a counter-orbiting one could also produce pericenter clustering of eTNOs, while the polar one leads to a wider spread of eTNO inclinations. 

\end{abstract}


\section{Introduction} 
\label{section:intro}
The three-body problem is one of the oldest problems in astrophysics. No general solution is possible due to its chaotic nature; however, important results can be obtained using perturbative approaches. For instance, when one of the objects is much farther from the other two objects (in a hierarchical configuration), one can treat the influence of the further object as a weak perturbation and discover interesting dynamical properties. \cite{kozai_secular_1962} and \cite{lidov_evolution_1962} developed the framework to study such three-body systems. It was found that when the inner binary (composed of the two closely separated objects) are inclined ($i_m>40^\circ$) with respect to the orbit of the farther object (outer orbit), Kozai-Lidov resonances could lead to large amplitude eccentricity and inclination variations of the inner binary.  

Due to the prevalence of three-body systems in hierarchical configurations, the Kozai-Lidov mechanism has since been applied to explain a wide variety of astrophysical phenomena. Inspired by the objects in our solar system, the Kozai-Lidov mechanism focused on systems with a circular outer orbit. In this case, expanding the disturbing function to second order in the small parameter (semi-major axis ratio of the inner and outer orbit, $a_1/a_2$) is sufficient. However, the quadrupole limit cannot describe the dynamics when the perturber is eccentric \citep[e.g.,][]{naoz_secular_2013, naoz_hot_2011-1, katz_long-term_2011}, and the disturbing function needs to be expanded up to the octupole order $(a_1/a_2)^3$. With an eccentric outer perturber, the eccentricity of the inner orbit can be excited close to unity. In addition, the inclination of the inner orbit can cross $90^\circ$, even when it starts in a near-coplanar configuration with respect to the outer binary \citep{li_eccentricity_2014}. This can explain a wide range of astrophysical phenomena (see \citealt{naoz_eccentric_2016} for a review).  

While the Kozai-Lidov mechanism and the eccentric Kozai-Lidov mechanism mainly describe the evolution of a test particle perturbed by an outer massive object, the evolution of an outer test particle perturbed by an inner binary has also been studied \citep[e.g.,][]{naoz_eccentric_2017, vinson_secular_2018, de_elia_inverse_2019}. In contrast to the inner test particle case, the quadrupole resonance allows the outer test particle's orbit to flip without changing its eccentricity. Higher order resonances can further excite test particle eccentricity and inclination. This has important implications for the dynamical evolution of debris disk surrounding planets in eccentric orbits \citep[e.g.,][]{Zanardi17}. 

Beyond the hierarchical limit, non-hierarchical dynamics also has wide applications but has been less explored. Recently, using a large ensemble of N-body simulations, \citealt{Stone19} obtained a statistical solution to the chaotic non-hierarchical three-body problem, under the assumption of ergodicity. They found that the non-hierarchical triple interactions almost always lead to a single escaping object and a stable bound binary. In addition, the eccentricity of the surviving binary follows a super thermal distribution. 

In mildly hierarchical triples, in which the triple system could still survive, non-secular effects can become important and can enhance the inner binary eccentricity \citep[e.g.,][]{cuk_secular_2004, antonini_secular_2012, antonini_black_2014}. In particular, \citealt{luo_double-averaging_2016} showed that short-time-scale oscillations can accumulate and make the secular results unreliable. They obtained `corrected double averaging' equations to account for the error in the secular results. Based on this correction, \citet{Grishin18} obtained analytical results on the maximum eccentricity of the inner binary at the quadruple level. 

However, the secular results can still provide a good approximation when the perturber is much less massive than the central object, and the corrected terms at the quadruple level are not sufficient when the three objects are closer to each other (with semi-major axis ratio $\gtrsim 0.3$). In particular, \citet{gronchi_averaging_1998} studied the dynamics of near earth asteroids, and developed a secular method without expansion of the semi-major axis ratio. More recently, this method has been used to study secular interactions between extreme TNOs and the hypothetical Planet-9 \citep{beust_orbital_2016, saillenfest17, li_secular_2018}. These studies mostly focused on the near-coplanar regime, where Planet-9 is located near the ecliptic plane. Nevertheless, a systematic study of dynamics of the mildly-hierarchical systems is missing in the literature, and the effects of an inclined Planet-9 on the extreme TNOs remains poorly explored.  

In this paper, we study the secular interactions in mildly-hierarchical systems with either an inner or an outer perturber. We compare the dynamical features to those in the hierarchical limit using the surface of section, which identifies the location of resonant and chaotic regions. We also identify initial conditions which can lead to large eccentricity and inclination variations. Finally, we apply our results to study interactions between objects in the outer solar system and the hypothetical Planet-9. The paper is organized as the following: in \textsection \ref{section:nummeth} we discuss the numerical techniques we adopt in this work, and in \textsection \ref{section:Val}, we derive the limit where the secular approach is valid. Then, we present the surface of sections in \textsection \ref{section:soc}, and the maximum eccentricity and inclination variations in \textsection \ref{section:emax}. In \textsection  \ref{section:outersol}, we apply our results to the objects in the outer solar system. Finally, we discuss our results and conclude in \textsection \ref{section:conc}. 

\section{Numerical Methods} 
\label{section:nummeth}

\begin{figure}[h]
\includegraphics[width=1.0\linewidth]{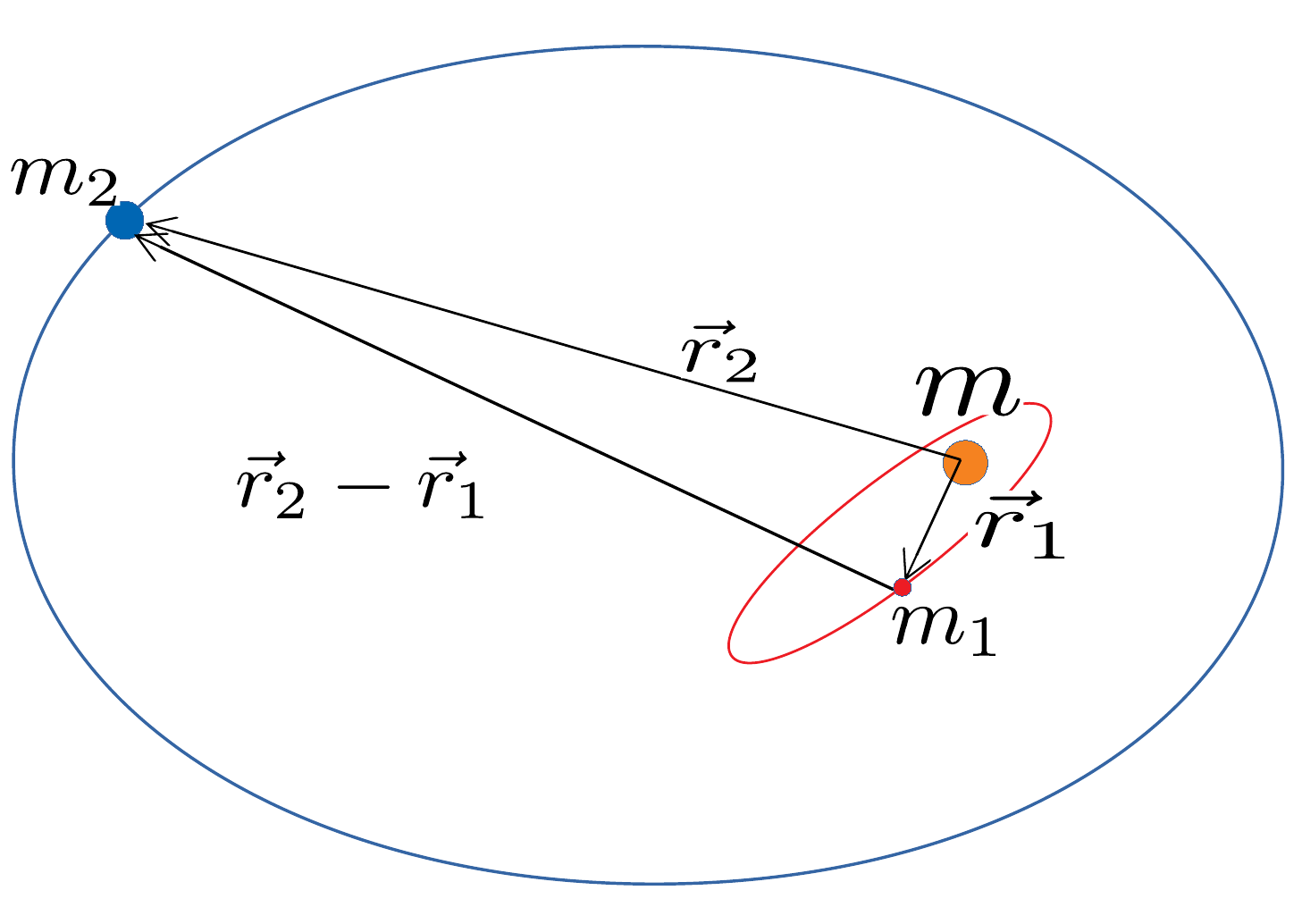}
\caption{Configuration of a mildly-hierarchical triple system. It comprises an inner binary, $m$ and $m_1$, and an outer binary formed by $m_2$ orbiting around the central object ($m$). Orbital elements of the inner binary are denoted by subscript 1, and those of the outer binary are denoted by subscript 2. 
}
\label{fig:nht}
\end{figure}

Mildly hierarchical triples can be separated into two binaries (as shown in Figure \ref{fig:nht}). The closely separated two objects, $m_1$ and $m$, form the inner binary, and the outer binary is composed of the farther object, $m_2$, orbiting around $m$. Note that this is different from the generic set up of hierarchical triples, where the outer binary is typically set to originate from the center of mass of the inner binary. This makes little difference here, since the perturbing mass that we consider must be much smaller than the central object for the secular results to be valid. 

Here, we use subscript 1 to denote the orbital elements of the inner orbit, and subscript 2 for the outer binary: $r_1$ and $r_2$ represent the position vectors of $m_1$ and $m_2$ from the central massive object $m$. In this paper we study the dynamics of a test particle in the triple system, and consider configurations with both inner and outer massive perturber. Thus, depending on the configuration, either $m_1$ or $m_2$ could be the test particle. We use parameter $\alpha$ to denote the ratio of the semi-major axis of the test particle with respect to the semi-major axis of the perturber. $\alpha<1$ for the inner test particle scenario, and $\alpha>1$ for  the outer test particle case. Please note that in this setup the choice of reference frame is arbitary. In test particle approximation, the orbital elements of the perturber are constant and can be used to define the frame of reference. In most of the simulations, we choose a coordinate system in which the inclination($i_2$), argument of pericenter($\omega_2$) and longitude of ascending node ($\Omega_2$) of the perturber are set to be 0.

To describe the motion of a test particle under the influence of the central object and a lower mass companion, we follow the expressions below \citep[e.g.,][]{murray_solar_2000}: 
\begin{align}
    &\ddot{\vec{r}}_i = \nabla_i V_{central} + \nabla_i R_i , \text{ where} \\ 
    &V_{central} = \frac{Gm}{|\vec{r_i}|}, \\
    &R_i = \frac{Gm_j}{|\vec{r_j}-\vec{r_i}|} -\frac{Gm_j(\vec{r}_i.\vec{r}_j)}{|\vec{r}_j|^3}.
\end{align}
$R_i$ is the disturbing function, $\vec{r}_i$ and $\vec{r}_j$ are the positions of the test-particle and the perturber with respect to the central object respectively, $m_j$ is the mass of the perturber, $m$ is the mass of the central object and $G$ is the universal gravitational constant.  The disturbing function describes the interaction potential between the test particle and the perturber, as the effect due to the influence of the perturber on the central body.  The first term of the disturbing function is called the direct term and the second one is called the indirect term. 
Indirect terms don't contribute to secular perturbations, as they average to zero \citep[e.g.,][]{murray_solar_1999}. We can write the secular disturbing function as
\begin{align} \label{eq:da}
    &R_{secular,i} \nonumber \\
    = &\frac{G m_j}{4\pi^2} \int \int \frac{1}{|\vec{r}_i - \vec{r}_j|} dM dM' ,
\end{align}
which averages out the fast, orbital timescale variations. In this study we do not include relativistic effects and focus only on point mass Newtonian interactions. We apply our results to outer solar system (\textsection \ref{section:outersol}) where relativistic corrections are not important.

In the mildly-hierarchical limit, the semi-major axis ratio is no longer a small parameter, so we numerically average the disturbing function instead of expanding it in the ratio of semi-major axes. This allows us to explore regimes which are non-hierarchical. Orbital elements of the test particle are evolved using Hamilton’s equations. We use the scale invariant Delaunay conjugate variables to define our phase space. They are composed of conjugate pairs $J_z-\Omega$ and $J-\omega$, where in the test particle limit $J=\sqrt{1-e^2}$ and $J_z=\sqrt{1-e^2}\cos{I}$. The equations of motion are given by:
\begin{eqnarray} \label{eq:eqm}
\dot{\omega} =\frac{1}{\sqrt{Gma}}\frac{\partial R}{\partial J} &,&  \dot{J} = -\frac{1}{\sqrt{Gma}}\frac{\partial R}{\partial \omega}, \nonumber \\
\dot{\Omega} =\frac{1}{\sqrt{Gma}}\frac{\partial R}{\partial J_z} &,& \dot{J_z} = - \frac{1}{\sqrt{Gma}}\frac{\partial R}{\partial \Omega}.
\end{eqnarray}

We solve the above set of equations numerically\footnote{The code is publicly available via a git repository hosted at \url{https://github.com/bhareeshg/gda3bd}.}. In non-hierarchical systems, the orbits of the perturber and the test particle can intersect. Although in the un-averaged system the perturber and the test particle may not actually collide, their may coincide at one or more points during the process of averaging.  
When they do, the disturbing function and its derivatives have singularities which can be difficult to evaluate. We use the CQUAD integration routine from the GNU Scientific Library(GSL) to perform these integrations. The equations of motion are evolved using GSL's implementation of the Runge-Kutta-Fehlberg (4, 5) method. 

Furthermore, \cite{Gronchi01} show that when orbits cross (or intersect) the derivatives of the disturbing function are not continuous across the crossing. Extra care needs to be taken when orbits cross. Whenever orbits cross during a single time-step, we abandon the current step and take a specific time-step to land exactly at the intersection. To find the intersection, we use the mutual nodal distance as the independent variable, which allows us to choose a step to reach the intersection (following \citealt{Henon82}). This is similar to the approach used by \cite{saillenfest_non-resonant_2017}. Instead of using implicit integration schemes, we use forward (and backward) discrete differentiation to calculate derivatives of the disturbing function from the positive (and negative) direction approaching the intersection separately. This provides the equation of motion prior (and post) the crossing.   

\begin{figure}
\centering
\includegraphics[width=1.0\linewidth,height=0.8\linewidth]{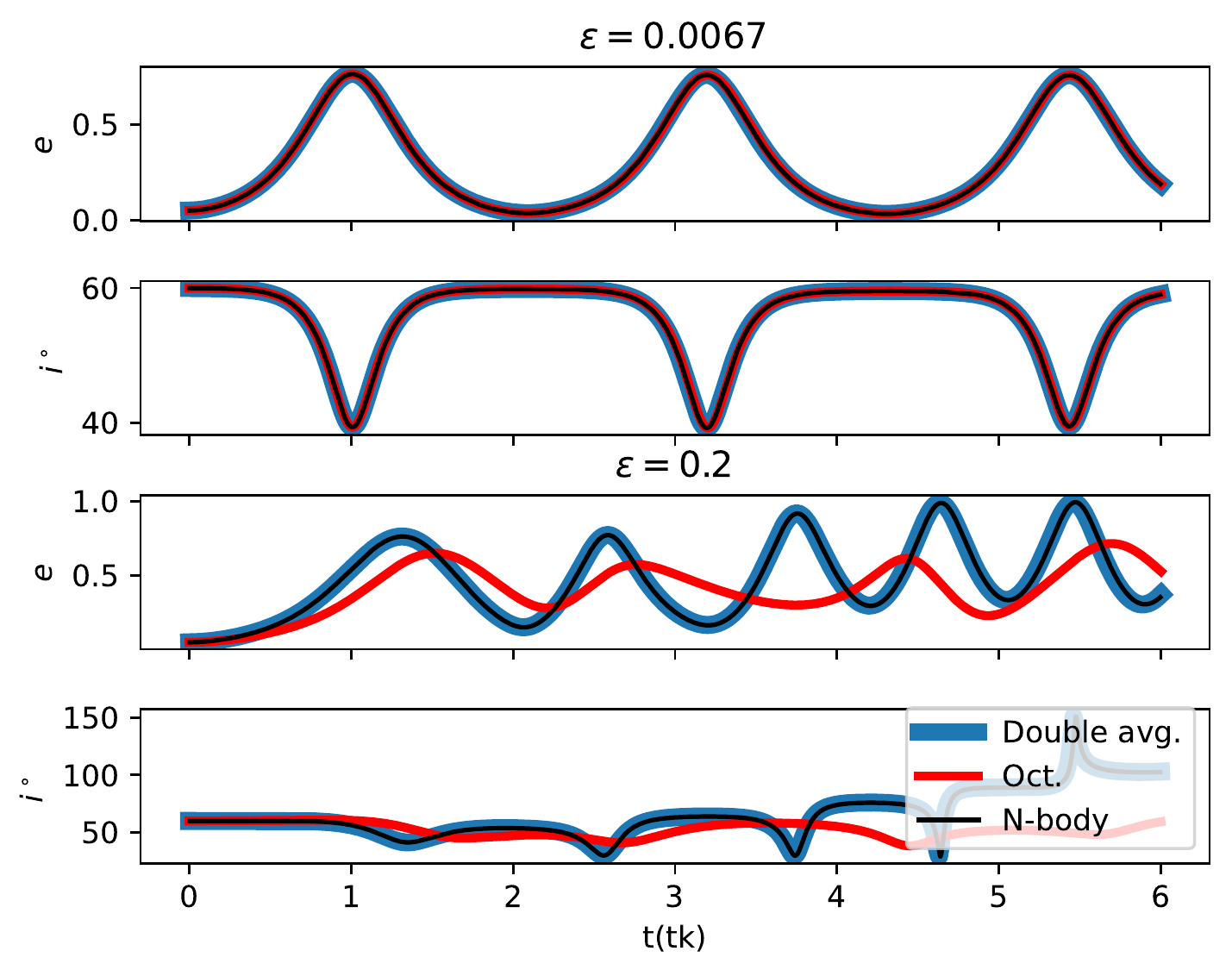}
\caption{Comparison with N-Body simulations. {\bf Top 2 panels}: Both octupole expansion and averaged disturbing function agree with N-body simulations for hierarchical systems. {\bf Bottom 2 panels}: For non-hierarchical systems, double averaged disturbing function agrees with N-body simulations but octupole expansion does not. The initial conditions for the upper panel are: $a_1=5\text{ AU}, e_1=0.05$, $i_1=60^\circ$, $\omega_1=0$, $\Omega_1=0,a_2=700\text{ AU}, e_2=0.6, i_2=0,  \omega_2=0,\Omega_2=0, m=1 M_\odot$ and $ m_2= 3 \times 10^{-5} M_\odot$. For the bottom panel we use the same initial conditions except $a_1=150\text{ AU}$.}
\label{fig:cmphvnh}
\end{figure}
In Figure \ref{fig:cmphvnh}, we show the orbital evolution of a test particle in a hierarchical system (top 2 panels) and in a mildly-hierarchical system (bottom 2 panels) as calculated by double averaged simulations (blue), octupole order simulations (red) and N-body simulations (black). We used the Burlisch-Stoer integration scheme from the  \texttt{MERCURY} simulation package \citep{chambers_mercury_1997} with a time step of $5\%$ of the period of the inner orbit to obtain the N-body results.  We show the time in terms of the Kozai timescale given by: 
\begin{equation}
	t_k=\frac{m}{m_{2}} (1-e_{2}^2)^{3/2} \frac{a_{2}^3}{a_1^3} P, \label{eq:tk}
\end{equation}
where $P$ is the orbtial period of the inner orbit \citep[e.g.,][]{valtonen_three-body_2006}. We use the parameter $\epsilon$ to quantify the level of hierarchy: 
\begin{equation}
    \epsilon=\frac{a_1}{a_2}\frac{e_2}{1-e_2^2}.
\end{equation}
The octupole order disturbing function can accurately describe triple system dynamics as long as $\epsilon<0.1$. In hierarchical systems (top), all three methods agree with each other. In non-hierarhical systems (bottom) on the other hand, the octupole order evolution does not agree with N-body simulations. Meanwhile, double averaged evolution is in excellent agreement with N-body simulations.

\subsection{Comparison to previous works}
Most studies in literature rely on the expansions of the disturbing function to study secular evolution in triple systems. Most commonly, the disturbing function is expanded in the eccentricities and inclinations of the interacting particles following the Lagrange-Laplace secular theory \citep{murray_solar_1999}. Such expansions have been used to study the interactions between the planets in systems with low eccentricities and inclinations. A special case where the absolute inclinations of the interacting particles are high but the relative inclination are small has also been studied\citep{boue_compact_2014}. In hierarchical systems inclination and eccentricity variations can be very high, but the separation between the inner and outer objects are large. Hence, the disturbing function is expanded in the ratio of semi-major axes \citep{kozai_secular_1962,harrington_dynamical_1968}. Expansions up to quadrupole, octupole and hexadecapole order terms have been explored in detail in literature\citep[e.g.,][]{yokoyama_orbits_2003, naoz_eccentric_2016}. However, these expansions cannot be used to study non-hierarchical systems where the semi-major axis ratios can be large and orbits can even cross. 

Beyond three body systems, the more general problem of understanding the evolution of large number of interacting particles has also been studied. For instance, \cite{fouvry_secular_2017} study the secular evolution of multiple stars orbiting a black hole using kinetic equations derived from BBGKY hierarchy. \cite{hamers_secular_2016} develop a method to study the evolution of hierarchical multiple systems composed of nested binaries. This method relies on the expansion of the Hamiltonian in terms of the binary separation ratios. Follow up studies have included the effects of flybys, instantaneous perturbations \citep{hamers_secular_2018} and orbit-averaging corrections \citep{hamers_secular_2020}.


Similar to this study, multiple works have used numerical averaging to calculate secular interactions between particles. In particular, \cite{touma_gausss_2009} developed the Gaussian ring algorithm, based on the analytical calculation of the orbit-averaged perturbing acceleration due to the force of a softened ring as well as a second average over the perturbed ring itself by numerical quadrature. Applying the Gaussian ring algorithm, \cite{nesvold_2016} modeled effects of self gravity in a debris disk by calculating the secular interactions of concentric rings. To include a careful treatment at the point of orbital intersection, our approach closely follows that of \cite{gronchi_averaging_1998} as mentioned above. More recently this approach has also been used to study the interactions between Planet-9 and extreme trans-neptunian objects \citep{beust_orbital_2016-1, saillenfest_non-resonant_2017}.

\section{Accuracy of the secular results for a circular, outer perturber}
\label{section:Val}

The double averaging method outlined in Section \ref{section:nummeth} is limited to low mass perturbers. As the mass of the perturber increases, non-secular oscillatory effects, such as mean motion resonances and evection resonances, become important \citep{luo_double-averaging_2016, Grishin18}, and the secular approach is no longer reliable. For instance, in the secular regime the mass of the perturber is just a scaling factor which would only change the dynamical timescale of the system (as shown in Equations \ref{eq:da} and \ref{eq:eqm}). This is not true when the mass of the perturber is high. 

\begin{figure*}
\centering
\includegraphics[width=1.0\linewidth]{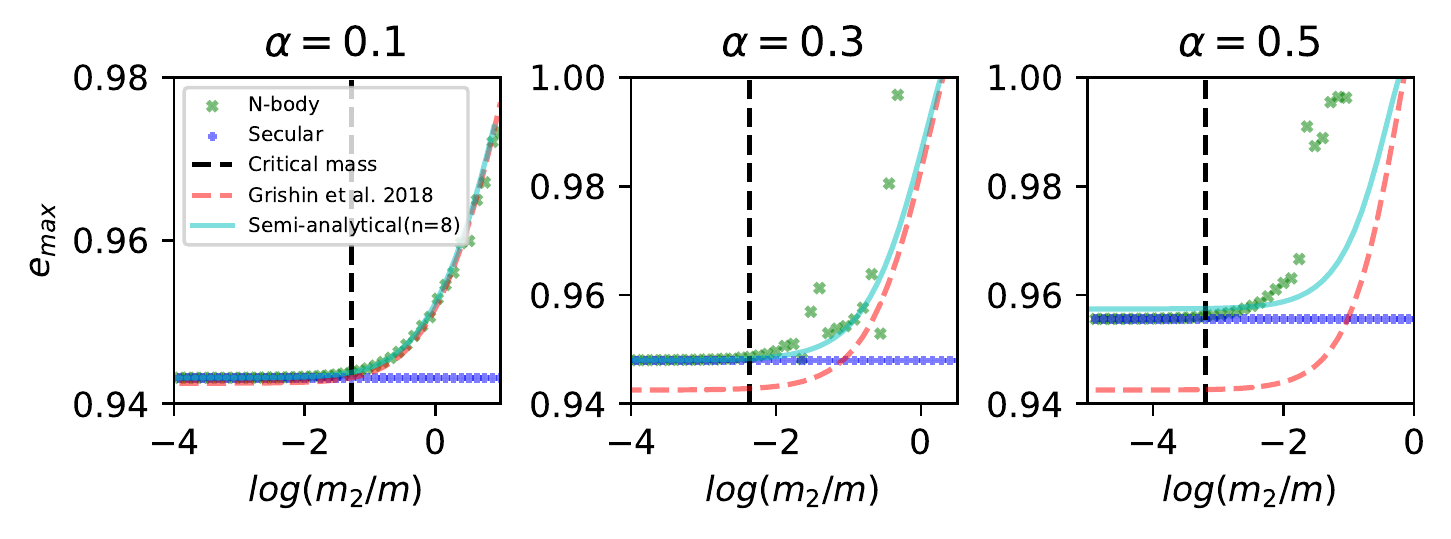} 
	\caption{Maximum eccentricity of the test particle as a function of the mass of a circular perturber. {\bf Left panel} corresponds to low $\alpha = 0.1$, {\bf middle panel} corresponds to $\alpha = 0.3$, and {\bf right panel} corresponds to $\alpha = 0.5$. The initial inclination($i_1$) and longitude of ascending node($\Omega_1$) of the test particle are  $75^\circ$ and $\pi/4$ respectively. Green crosses represent results using N-body simulations, and blue dots represent results using the double averaged secular simulations. The critical perturber masses beyond which the secular results deviate from that of N-body simulations are indicated by the black dashed line. As $\alpha$ increases, the critical mass decreases. Red dashed lines show $e_{max}$ obtained by \citet{Grishin18} including the single average oscillations, and cyan solid lines represent our results of $e_{emax}$, calculated using a higher order semi-analytical method (using terms of disturbing function up to $\alpha^8$). This method provides more accurate estimates of $e_{max}$ at higher values of $\alpha$. \label{fig:emaxcomp}}
\end{figure*}

To illustrate this, we perform an ensemble of double averaged secular simulations and compare them with N-body simulations. We focus on the case of an inner test particle under the perturbation from an object on a circular orbit. We run these simulations for three Kozai timescales (see Equation \ref{eq:tk}). To check for consistency, our N-body results were verified with two different integration methods: Burlisch-Stoer and hybrid integrators of the \texttt{MERCURY} package. In addition, we ran N-body simulations and  secular simulations with different time steps (2.5\%, 5\% and 10\% of the inner orbital period). The fractional differences in $e_{max}$ of the hybrid integrator do not converge for the different time steps when the eccentricity is high. Thus, we use the Burlisch-Stoer integrator for our N-body simulations. We find that the results of the secular method and the Burlisch-Stoer integrator are consistent across different choices of time steps, and the secular results are in excellent agreement with the N-body results (except near mean-motion resonances). The choice of time step did not affect our results as long as the time-step for N-body simulations computed using Burlisch-Stoer integrator was less than $\sim 10\%$ of the period of the inner most object and for secular simulations was less than $0.5\%$ of the secular timescale.

In Figure \ref{fig:emaxcomp}, we plot the maximum eccentricities against the mass of the perturber. The perturber is on a circular orbit, and test particles start on a circular orbit ($e_1=0$) with $i_1=75^\circ$ and $\Omega_1=\pi/4$. Three different configurations of the test particle are shown: the panel on the left has a semi-major ratio of $\alpha=0.1$, the one in the middle corresponds to $\alpha=0.3$, and the one on the right corresponds to $\alpha=0.5$. The blue pluses represent the double average results, and the green crosses represent the N-body results. We note that the system becomes chaotic when it gets less hierarchical ($\alpha \gtrsim 0.18$). While the qualitative behavior can still be valid, chaos could affect the specific value of $e_{max}$ for a certain configuration.

The maximum eccentricity obtained from the double average results is independent of the perturber's mass, because the masses only determine the dynamical timescale of the system. For the n-body results, as the mass of the perturber increases so do the non-secular effects, causing $e_{max}$ to increase. A purely secular double averaging method fails to capture this effect. Thus, as the perturber become more massive, the maximum eccentricity calculated from the double averaged method ($e^{DA}_{max}$) starts to deviate from that calculated using N-body simulations ($e^{NB}_{max}$), as shown in Figure \ref{fig:emaxcomp}. 

We use black dashed lines to denote the critical perturber mass, the mass of the perturber above which the double averaged results start to deviate from the N-body results. Specifically, the critical mass is defined by the following criteria:
\begin{align}
\frac{|(1 - e^{NB}_{max}(m_2)) - (1 - e^{DA}_{max})|}{(1 - e^{DA}_{max})} > 0.01.
\label{eq:mcrit}
\end{align}
As the semi-major axis ratio increases, the non-secular effects of the perturber becomes stronger and the critical mass decreases. 

For low values of $\alpha$, the deviation from the secular results can be explained by single averaged corrections, which are obtained by averaging the disturbing function only over the orbit of the inner test particle (the shortest time scale in the problem). \citet{luo_double-averaging_2016} derived single averaged corrections to the double averaged disturbing function. Using their derivation and the quadrupole order double averaged disturbing function, \cite{Grishin18} derived single averaged correction ($\delta e^{SA}_{corr}(m_2,\alpha)$) to $e^{DA}_{max}$. In Figure \ref{fig:emaxcomp}, red dashed lines represent the results by \citet{Grishin18}. While the quadrupole result agree very well with N-body results at $\alpha=0.1$, at higher values of $\alpha$, it fails to agree even for low $m_2$.  

Therefore, we use a semi-analytical method which includes terms up to $\alpha^8$ to calculate $\bar{e}^{SA}_{max}$, while using the same single averaged correction ($\delta e^{SA}_{corr}(m_2, \alpha)$) from \citet{Grishin18}. We focus on an object in the resonant region following  a librating trajectory. Thus, $\omega_1 = \pi/2$ when $e_1=0$ and $e_1=e_{max}$. 
Since $j_z$ is constant under the perturbation of objects in circular orbits, we obtain the inclination when the eccentricity assumes it's maximum value: 
\begin{align}
    i_{emax}=\arccos(\cos(i_0)/\sqrt{1-(\bar{e}^{SA}_{max})^2}).
\end{align}
When the orbit of the perturber is circular, the Hamiltonian is independent of $\Omega$, and only terms of even order survive in the Hamiltonian in addition to the the single averaged correction:
\begin{align}
H=H_{n=2} + H_{n=4} + H_{n=6} + H_{n=8}+H_{SA} ,
\end{align}
where the single averaged correction is derived in Eqn. (39) of \citet{luo_double-averaging_2016}:
\begin{align}
    &H_{SA}=-\frac{27}{8}\epsilon_{SA}j_z(\frac{1-j^2}{3}+3e^2+5e^2\cos(i)^2) ,\nonumber \\
    &\epsilon_{SA} = \Big(\frac{a_1}{a_2(1-e_2^2)}\Big)^{3/2}\Big(\frac{m_2^2}{(m+m_2)m}\Big)^{1/2}.
\end{align}
In this case, the maximum eccentricity can be obtained numerically, requiring that energy is conserved: 
\begin{align}
	&H(e=0,i=i_0,\omega=\pi/2) \nonumber \\
	= &H(e=\bar{e}^{SA}_{max},i=i_{emax},\omega=\pi/2)  .
\end{align}

We show the semi-analytical results (cyan lines) in Figure \ref{fig:emaxcomp}. By including the $8^{th}$ order, the semi-analytical approach agrees very well with the direct N-body simulations at higher values of $\alpha$ and low values of $m_2$. However, the semi-analytical results deviate from the N-body results for larger $m_2$ when $\alpha \sim 0.5$, suggesting higher order single averaged corrections need to be included. Please note that the approach outlined above for the 8th order expansion works only for a circular perturber. When the perturber is eccentric, the number of degrees of freedom for the system increases to 2 and $e_{max}$ cannot be easily obtained. In addition, wiggles in the numerical results are mainly due to close encounters, which cannot be captured by the secular approach. 

In the end, we derive an analytical expression for the critical perturber mass, below which the double average provides a good approximation, using the results of the maximum eccentricities. 
When the perturber is on a circular orbit, $j_z$ is a constant of motion as a result of the double average over the inner and outer binary orbits. However, single averaged corrections lead to oscillations in $j_z$ with maximum fluctuations proportional to $\epsilon_{SA}$. Assuming all the deviations between $e^{DA}_{max}$ and $e^{NB}_{max}$ are due to fluctuations in $j_z$ and are caused by the single average oscillations, we can obtain the fractional change in the maximum eccentricity due to the single averaged oscillation effects as the following:
\begin{align}
	f_e = \frac{\delta e^{SA}_{corr}(m_{2},\alpha)}{1 - {e}^{DA}_{max}(\alpha)} ,
\end{align}
where $f_e = 0.01$ gives the critical mass of $m_2$ according to our definition in Eqn. (\ref{eq:mcrit}), and $e^{SA}_{corr}$ and $e^{SA}_{max}$ in the quadrupole limit are obtained in \citet{Grishin18} (Eqns. 33 and 38):
\begin{align}
	\delta e^{SA}_{corr} = &\frac{135}{128}\bar{e}_{max}^{SA}\epsilon_{SA}\Big(\frac{16}{9}\sqrt{\frac{3}{5}}\sqrt{1-(\bar{e}^{SA}_{max})^2} \nonumber \\
	&+\epsilon_{SA}-2\epsilon_{SA}(\bar{e}^{SA}_{max})^2\Big), \\
	e^{SA}_{max} = & \sqrt{1-\frac{5}{3}\cos^2{i_0}\frac{1+\frac{9}{8}\epsilon_{SA}\cos{i_0}}{1-\frac{9}{8}\epsilon_{SA}\cos{i_0}}}.
	\label{eqn:fe}
\end{align}
To the first order in $\epsilon_{SA}$ and assuming $m_2 \ll m$, we can get an analytical expression for the critical mass $m_{2, crit}$, where $f_e<C$:
\begin{align}
    m_{2, crit} = \frac{4C m (6 - \sqrt{6 - 30 \cos{(2i_0)}})}{15\alpha^{3/2}\sqrt{
 \cos^2(i_0)(9 - 15 \cos^2(i_0))}}.
 \label{eq:mccri}
\end{align}

\begin{figure}[h]
\centering
\label{fig:alphamass}
\includegraphics[width=1.0\linewidth]{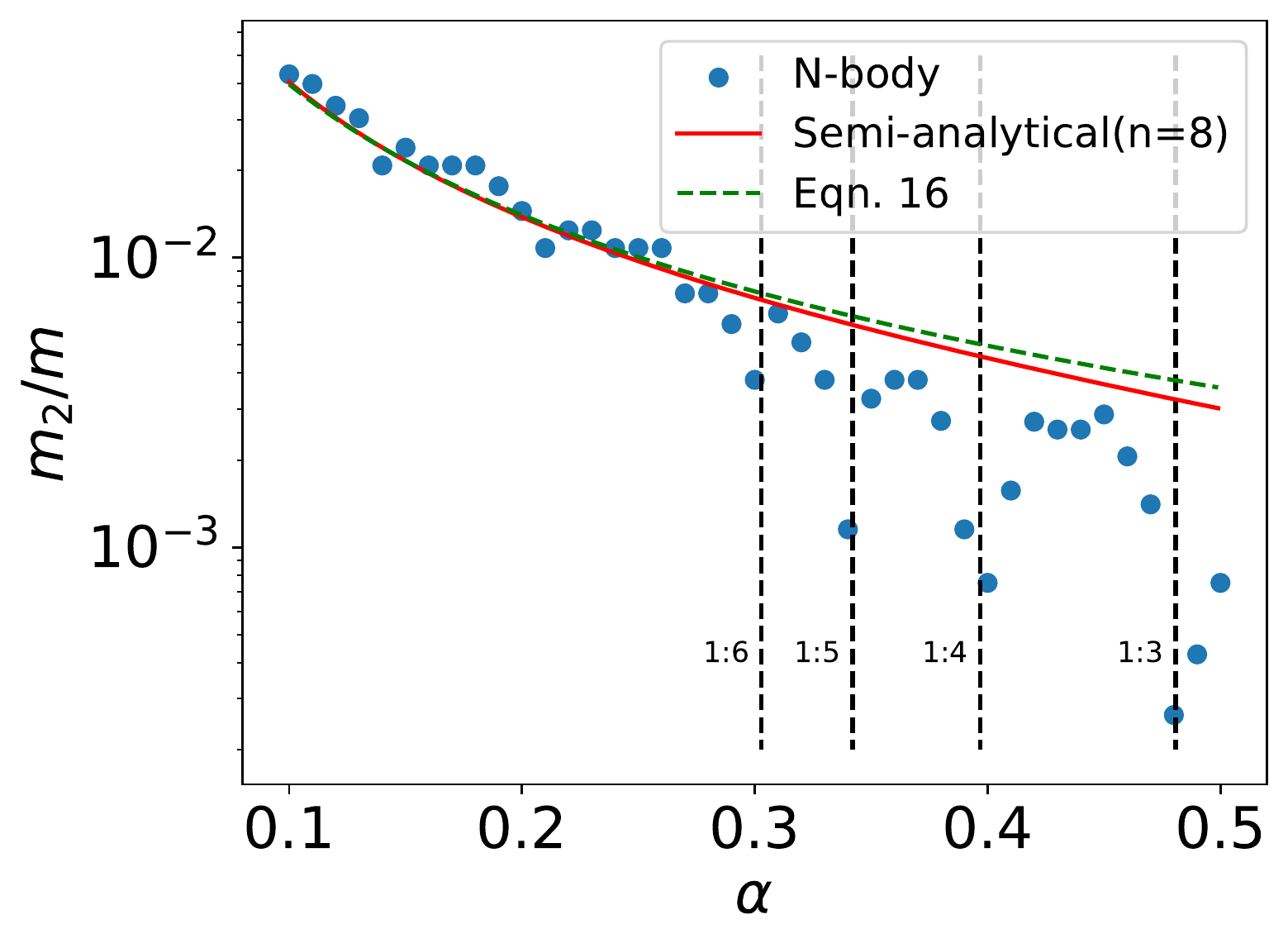}
\caption{Critical mass as a function of the ratio of semi-major axes ($\alpha$). The blue dots show the results using N-body simulations. We define critical mass as the minimum mass of the perturber above which the maximum eccentricity derived using secular approximation differs from that derived from N-body simulations (Equation \ref{eq:mcrit}). As the semi-major axis ratio ($\alpha$) increases, critical mass decreases. We derived an analytical expression for critical mass (green dashed line, Equation \ref{eq:mccri}), which agrees well with the N-body results when $\alpha \lesssim 0.3$. The red solid line represents the semi-analytical results up to the $8^{th}$ order in semi-major axis ratio, and it agrees slightly better than the analytical results. The dashed black lines represent the location of the mean motion resonances.}
\end{figure}

We compare the expression with N-body simulations in Figure \ref{fig:alphamass}, which shows the critical mass as a function of $\alpha$. The dots represent the N-body results, the dashed line represents the analytical results in Eqn. \ref{eq:mccri}, and the solid red line represents the semi-analytical results based on the $8^{th}$ order expansion in semi-major axis ratio ($(a_{in}/a_{out})^8$). When the perturber is close to the test particle $a_{1}/a_{2}\sim0.5$, the perturber needs to be small comparing with the central object ($\sim 10^{-3.5}$ that of the central object's mass). When the perturber is farther ($a_{1}/a_{2}\sim0.1$), the perturber only needs to be lower than $\sim 10^{-1.5}$ times the mass of the central object for the double average to be valid. 

Our analytical expression agrees well with the N-body results when $\alpha \lesssim 0.3$. When $\alpha \gtrsim 0.3$, effects of mean motion resonances ($P_t:P_2 = 1:5, 1:4, 1:3$) become important, and this makes the analytical results based on the single averaged method no longer reliable. The semi-analytical results up to the $8^{th}$ power in $\alpha$ agree slightly better compared to the analytical expression. When $\alpha \gtrsim 0.3$, the single averaged correction becomes similar in magnitude to the expansion at the $n=8$ order (up to $(a_{1}/a_{2})^8$). Thus, one needs even higher order expansions to obtain more accurate semi-analytical results on the critical mass. 

\section{Surface of sections} 
\label{section:soc}

For systems with perturbing mass much lower than that of the central object($\sim 10^{-3}$ to $10^{-2} m$), the secular results using the double average method provide a good approximation to the dynamics.  
Thus, to better understand the dynamics of mildly hierarchical triples, we analyze the secular results in this section in more detail. In the test particle limit, the secular dynamics can be reduced to two degrees of freedom, and the phase space is four dimensional. It is difficult to visualize the phase space directly and thus, we use surface of sections to characterize the dynamical properties, similar to the approach in e.g., \citealt{li_chaos_2014}.

We first look at surface of sections in $e\cos\omega-e\sin\omega$ space. We perform an ensemble of secular simulations using the method outlined in Section \ref{section:nummeth}. From the trajectories of the test particles evolved in these simulations, we collect all points on the surface $\Omega=0$ for inner test particle configurations and $\Omega=\pi/2$ for outer test particle configurations which satisfy the condition $\dot{\Omega}<0$. We choose $\Omega=\pi/2$ for the outer test particle configurations because the librating trajectories are centered around $\Omega=\pi/2$ due to the quadrupole resonances \citep{naoz_eccentric_2017}.

To study how the surfaces change as the system become less hierarchical, we make surfaces for different values of the semi-major axes ratio ($\alpha =\{0.1,0.2,0.3,0.5,2,3,5\}$), and we also consider perturbers with different eccentricities ($e_2=\{0.2,0.4,0.6,0.8\}$).  We show a few representative panels in Figure \ref{fig:tsoce}. See the Appendix for the full set of surfaces. We take the argument of pericenter ($\omega_2$) and longitude of ascending node($\Omega_2$) of the perturber to be zero. 


\begin{figure}[h]
\centering
\includegraphics[width=1.0\linewidth]{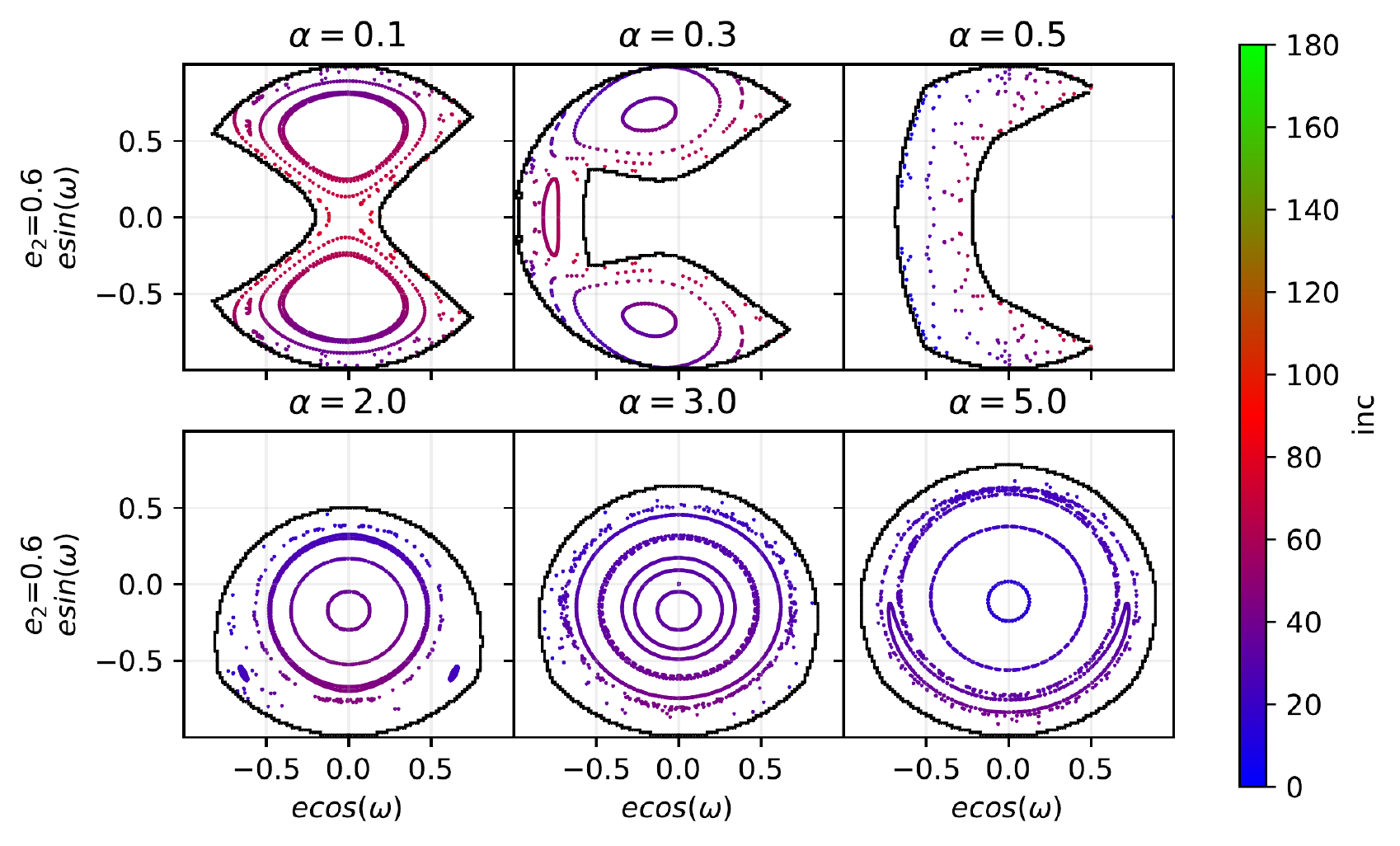}
\caption{Surface of section in $e\cos(\omega)-e\sin(\omega)$ space for $e_2=0.6$. We use the conditions $\Omega=0$ and $\dot{\Omega}<0$ for inner test particle configurations and $\Omega=\pi/2$ and $\dot{\Omega}<0$ for outer test particle configurations to choose our points. \label{fig:tsoce}}
\end{figure}

As the systems become less hierarchical, physically allowed regions change, particularly for those with an eccentric perturbers $e_2\gtrsim 0.4$. For instance, regions inside $-\pi/2 < \omega <\pi/2$ can be allowed in the mildly hierarchical configurations with an eccentric perturber, as shown in the top right panel in Figure \ref{fig:tsoce}. However, the dynamical features typically have a weak dependence on the semi-major axis ratio in the same dynamically allowed regions. Similar to the hierarchical limit, where resonances have been identified at $\omega=\pi/2 \text{ and }3\pi/2$ \citep{kozai_secular_1962, lidov_evolution_1962} in the quadrupole limit, and at $\omega=\pi$ in the octupole limit in surfaces \citep{li_chaos_2014}, we identify resonances at $\omega=\{\pi/2, \pi, 3\pi/2\}$ in the inner test particle case. Overall, the system becomes more chaotic when it is less hierarchical.

For outer test particle configurations (bottom row of Figure \ref{fig:tsoce}), we can see resonances at $\omega=3\pi/2$ (note that resonances at $\omega = \pi/2$ can also be seen in some surfaces, e.g., middle panel in $e_2=0.6$, $\alpha=5$ in the Appendix). This is also similar to the hierarchical limit. Specifically, $\omega$ has no resonances at the quadrupole limit, but there are resonances at $\omega=\pi/2$ in the octupole order Hamiltonian \citep{naoz_eccentric_2017}. Note that \citet{naoz_eccentric_2017} identified resonances at $\pi/2$, because they use $\dot{\Omega}>0$ as their condition to make the surfaces. To focus on the low inclination dynamics, we use $\dot{\Omega}<0$ instead, since the inclination of points with $\dot{\Omega}>0$ are typically over $90^\circ$. Similar to the inner test particle cases, as the system become less hierarchical, there are generally more chaotic regions at higher values of $e_2$. The chaotic regions result from overlapping of resonances, and we can see higher order resonances embedded in chaotic regions (See the panel for $\alpha=2$ , $e_2=0.2$). 

The colors in the surfaces represent inclination of the particles, which oscillate as the eccentricities vary. For example, trajectories librating around $\omega=\{\pi/2,3\pi/2\}$ undergo inclination variations that can flip over $90^\circ$. Different from the hierarchical limit, orbits can flip from a near coplanar configuration in a near circular orbit under the influence of an eccentric outer perturber. We will discuss this in more detail in Section \ref{section:emax}. 

We now look at surface of sections in $i\cos(\Omega)-i\sin(\Omega)$. We plot surface of sections for systems with the same set of configurations as discussed above. In Figure \ref{fig:tsoci} we show surfaces for $\alpha=\{0.1,0.3,0.5,2,3,5\}$ and $e_2=\{0.6\}$ (see the Appendix for the rest of surfaces). We chose the surface $\omega=0$ with the condition $\dot{\omega}>0$ to make these plots. Color in this case represents the eccentricities of the test particles. Similar to the surfaces in $e\cos(\omega)-e\sin(\omega)$, physically allowed regions change for mildly hierarchical configurations with highly eccentric outer perturbers. In particular, additional resonant regions show up around $\Omega=0$ with high inclinations and moderate eccentricities (e.g., top right panel in Figure \ref{fig:tsoci}), under perturbations of an eccentric outer object ($e_2 \gtrsim 0.4$) with $\alpha \gtrsim 0.3$.

\begin{figure}[h]
\centering
\includegraphics[width=1.0\linewidth]{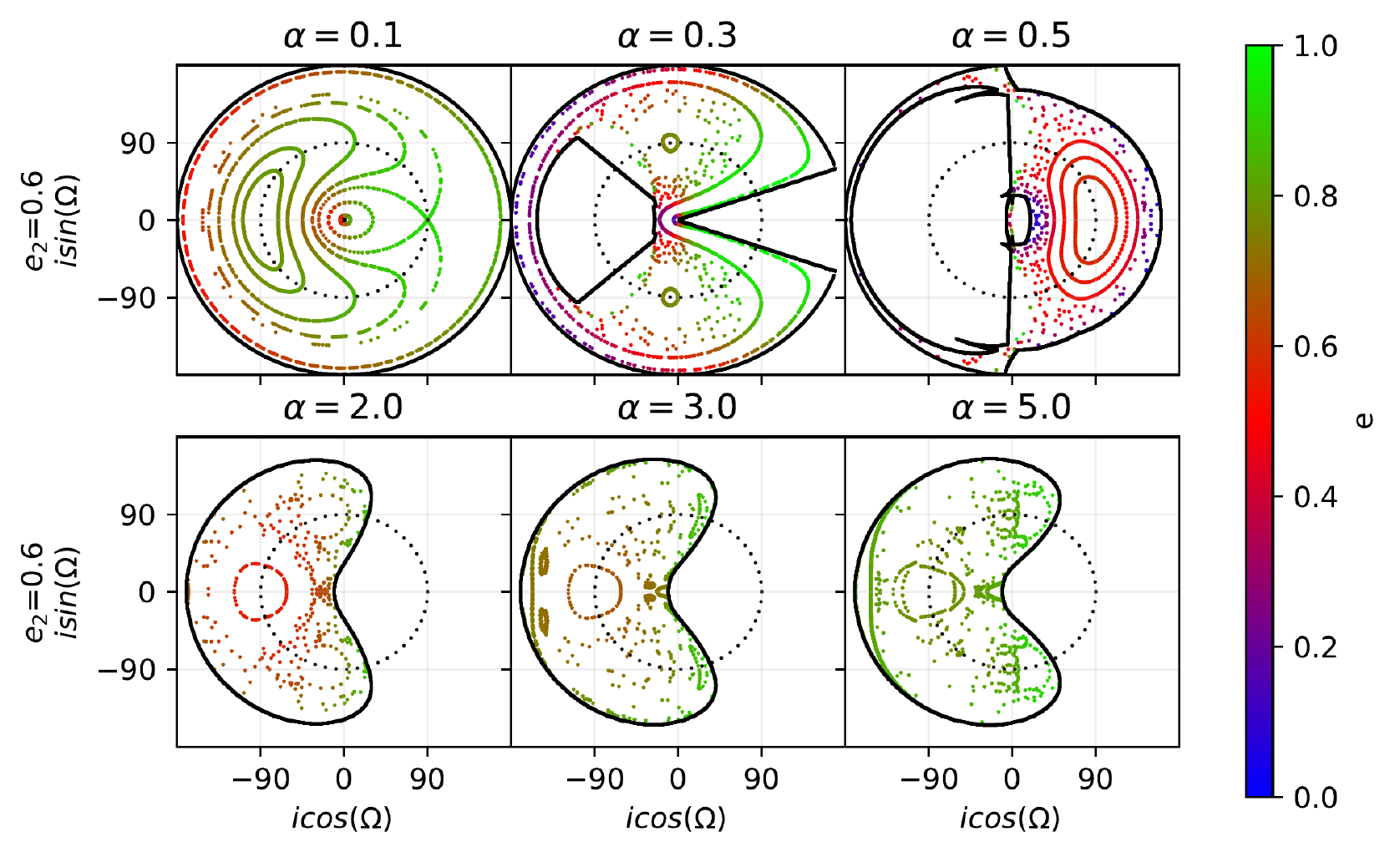}
\caption{Surface of section in $i\cos(\Omega)-i\sin(\Omega)$ space for $e_2=0.6$. We use the conditions $\omega=0$ and $\dot{\omega}>0$ to choose our points.  Dashed circle in black corresponds to an inclination of 90$^\circ$. \label{fig:tsoci} }
\end{figure}

At low values of $\alpha$, resonances around $\Omega=\pi$ lead to high eccentricity excitation as the orbits flip cross $90^\circ$. In addition, test particle's orbits circulate when the inclination is close to $0^\circ$ and $180^\circ$ due to an eccentric perturber (e.g., top left panel of Figure \ref{fig:tsoci}). The orbits librate for a wide range of inclinations in between the coplanar and counter orbiting configurations. To illustrate the flips of the orbit, we marked the circle with a radius of $90^\circ$. Many librating orbits cross over $90^\circ$ at high eccentricities, similar to the hierarchical limit. When the systems become less hierarchical, eccentricity typically oscillates faster than inclination, and remains low when $\omega$=0. Thus, we find particles with moderate or low eccentricities near $90^\circ$ on the surfaces (e.g., top right panel of Figure \ref{fig:tsoci}). However, the orbits still flip at relatively high eccentricities ($e>0.8$), and the eccentricity when the orbit flips do not show a clear trend as a function of $\alpha$.





The resonant regions in the plane of $i\cos{\Omega}-i\sin{\Omega}$ are again similar to that in the hierarchical limit. We identify resonances at $\Omega=\{0, \pi, \pi/2$ and $3\pi/2\}$ in inner test particle configurations. Note that the resonant regions around $\pi/2$ and $3\pi/2$ exist in the surface of e.g., $\alpha = 0.3$, $e_2=0.2$ near $i\sim 90^\circ$ (as shown in the the fifth column in Figure \ref{fig:i2} in the Appendix). This is similar to the octupole order, where \cite{li_chaos_2014} find resonances at $\Omega=\{0, \pi\}$. We can also identify higher order resonances with high $(\alpha)$ values embedded in the chaotic sea. Similar to the surfaces in $e\cos{\omega}-e\sin{\omega}$, chaotic regions are more common at higher $\alpha$ values. For outer test particle configurations, we can identify resonances at $\Omega=\{\pi/2,\pi,3\pi/2\}$. This is consistent with the hierarchical limit where at the quadrupole order, $\Omega$ is the resonant angle which librates around $\Omega=\pi/2$ \citep{naoz_eccentric_2017}. Similar to the surfaces shown in $e\cos{\omega}-e\sin{\omega}$, the size of the chaotic regions increases as the systems become less hierarchical.

Different from the octupole level of approximation, the physical regions change when the system becomes less hierarchical. In particular, the region with $90^\circ<\Omega<270^\circ$ becomes largely un-physical (e.g., with $\alpha=\{0.3\}, e_2=\{0.6,0.8\}$, and with $\alpha=\{0.5\},e_2=\{0.2,0.4,0.6, 0.8\}$). However, the dynamical features are still analogous to the octupole cases, as mentioned above. In particular, in mildly-hierarchical systems (e.g., with $\alpha=0.1, e_2={0.8}$, with $\alpha=\{0.3\}, e_2=\{0.4, 0.6,0.8\}$, and with $\alpha=\{0.5\},e_2=\{0.2,0.4,0.6, 0.8\}$), libration regions appear near $i=90^\circ$ with $-90^\circ < \Omega < 90^\circ$. The location of these higher order resonances changes with the value of the energy. At low energies (e.g., in the middle upper panel of Figure \ref{fig:tsoci}), these libration regions occur near $\Omega=\pi/2$ and $3\pi/2$. As the energy increases, these two libration regions move towards $\Omega=0$. At some point they overlap and lead to resonances at $\Omega=0$ (e.g., in the upper right panel of Figure \ref{fig:tsoci}). 

\section{Eccentricity and Inclination Excitation} 
\label{section:emax}

\begin{figure}[h]
\centering
\includegraphics[width=1.0\linewidth]{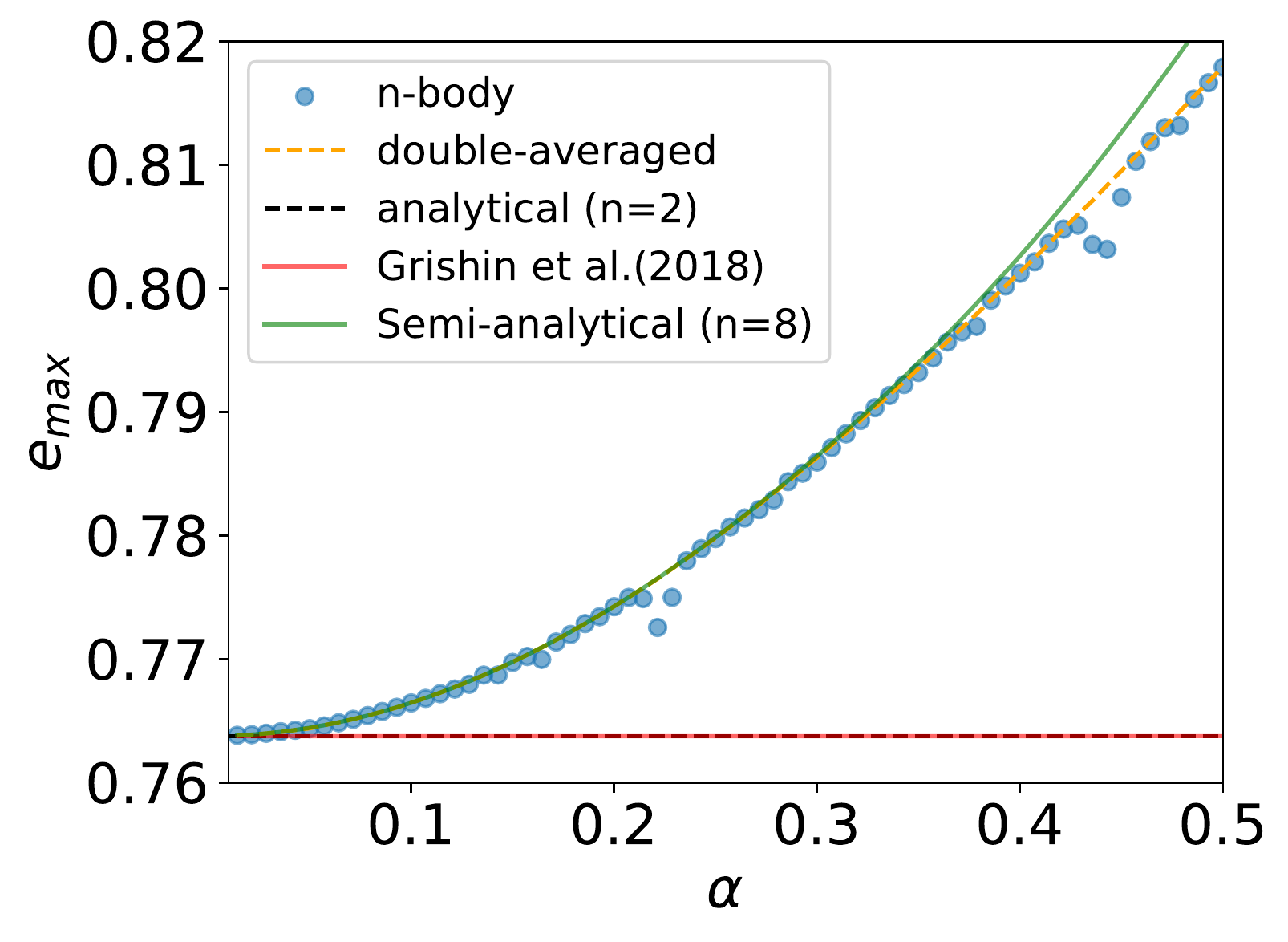}
\caption{Maximum eccentricity as a function of the semi-major axis ratio ($\alpha$) for the inner test particle configuration with a circular perturber. The test particle is initially on a circular orbit with $i =60 ^\circ, \omega=90 ^\circ$ and $\Omega=0$.  Results from N-body simulations are shown as blue dots whereas the results from secular simulations are shown as yellow dashed lines. It can be seen that results from secular simulations agree very well with N-body simulations for all values of $\alpha$. The quadruple level result is shown for comparison. We also show results from the semi-analytical method which uses terms in the disturbing function up to the order $\alpha^8$. While it agrees with N-body simulation till $\alpha=0.35$, at higher values of $\alpha$ this method over predicts $e_{max}$. \label{fig:kozaiemax}}
\end{figure}

The surface of section plots in Section \ref{section:soc} demonstrate that the eccentricity and inclination of the test particles can have large amplitude variations, and this may have important implications for astrophysical systems. 
In this section, we consider the eccentricity and inclination excitation in detail. First, to illustrate how the eccentricity excitation changes as the system become less hierarchical, we show the maximum eccentricity of the test particle as a function of the semi-major axes ratio ($\alpha$) in Figure \ref{fig:kozaiemax} for a circular perturber with a mass ratio $m_2/m=3 \times10^{-4}$. For these simulations we choose an initial inclination of $60^\circ$, greater than the minimum inclination$(~40^\circ)$ needed for the Kozai-Lidov mechanism to operate.

Figure \ref{fig:kozaiemax} shows that as the semi-major axis ratio increases so does the maximum eccentricity. While the quadruple order expression for the maximum eccentricity ($e_{max}=\sqrt{1-\frac{5}{3}\cos^2i_0}$) is accurate only for $\alpha<0.05$, the semi-analytical results at the $8^{th}$ order of expansion outlined in Section \ref{section:Val} agree more closely with N-body simulations for higher values of $\alpha$. For $\alpha>0.3$, this approach also ceases to agree with N-body results and predicts a higher value for maximum eccentricity. This indicates the importance of even higher order terms of the disturbing function in this regime. The double average results agree well with the N-body results with the low mass perturber. With a circular perturber, the maximum eccentricity can be quite large, but it is mostly well below unity. 


When the orbit of the perturber becomes eccentric, maximum eccentricity of the test particle becomes more complicated. Previous studies on hierarchical systems at the octupole order showed that the eccentricity of the test particle can be excited to values close to unity, and inclination can be excited to values beyond 90$^\circ$ \citep{li_eccentricity_2014,naoz_hot_2011,lithwick_eccentric_2011,katz_long-term_2011}. Orbits can be flipped (inclination crosses over $90^\circ$) in three different scenarios 1) low initial eccentricity and high initial inclination($i>40^\circ$ and $i<140^\circ$) a scenario similar to the standard Kozai-Lidov mechanism. 2) high initial ecentricity and low inclination and 3) medium eccentricity and high inclination. When the perturber's orbit is circular, the dynamics is mainly dominated by quadrupole resonances (as discussed in \textsection \ref{section:Val}), while when the orbit of the perturber is eccentric, both octupole and quadrupole resonances become important (See e.g., \citealt{li_chaos_2014}). 
\begin{figure}[h]
\centering
\includegraphics[width=1.0\linewidth]{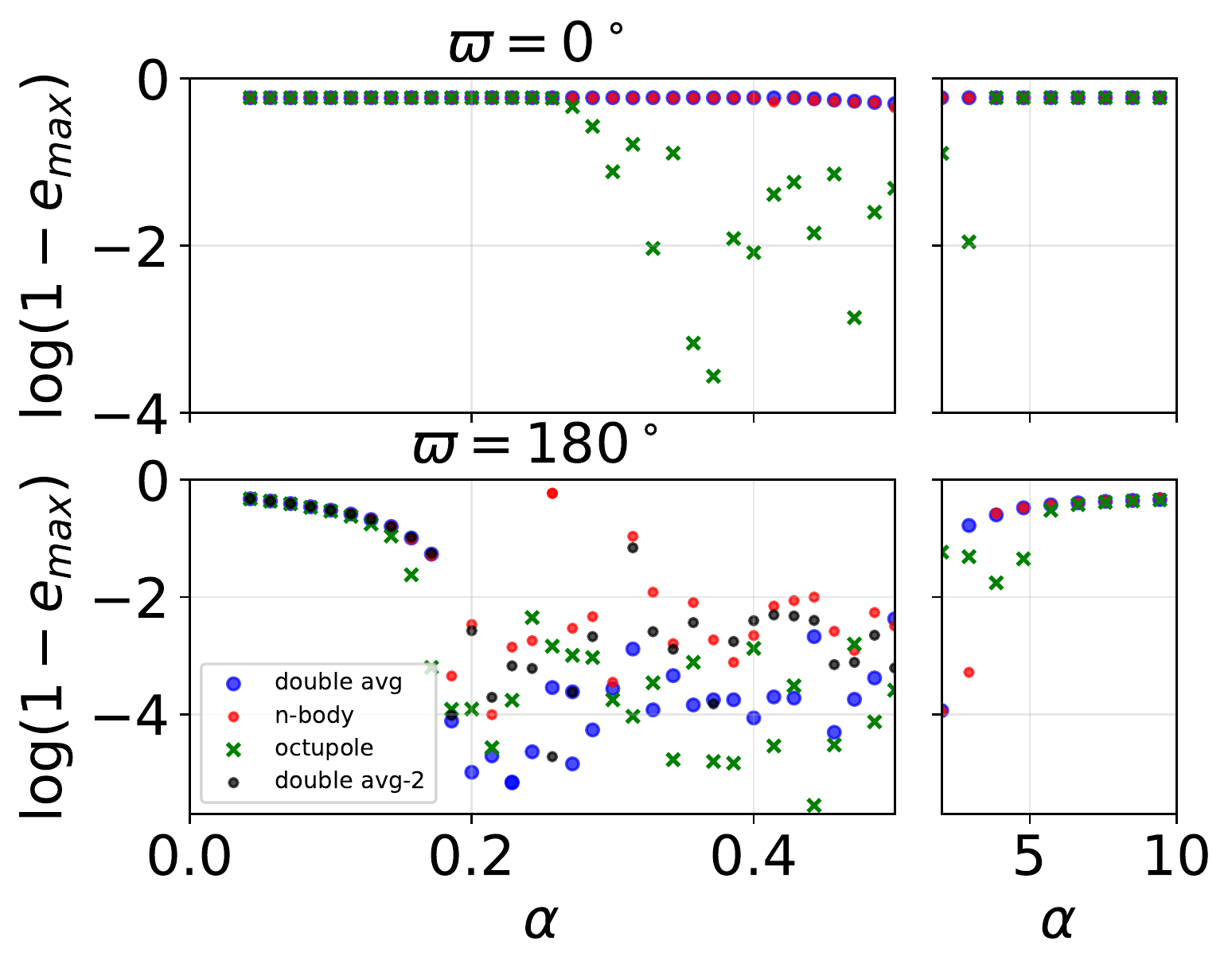}
\caption{Maximum eccentricity of {\bf initially nearly co-planar} test particles as a function of the semi-major axis ratio ($\alpha$) for the inner and the outer test particle configurations with an eccentric perturber. The {\bf top panels} include particles starting aligned with respect to the perturber, and the {\bf bottom panels} include those starting anti-aligned with respect to the perturber. Red dots represent results from N-body simulations, blue dots represent secular double averaged results and green crosses represent octupole results. To illustrate the effects of chaos, we use grey dots to represent the results from secular double average with a slightly different initial condition (``double avg-2''). The double averaged results are consistent with N-body results across all values of $\alpha$, except when the eccentricity is close to 1, and also at a few points where the particles are trapped in mean motion resonances. Octupole results agrees with N-body results only when $\alpha<0.1$ for the anti-aligned configuration and when $\alpha<0.25$ for the aligned configuration. For initially anti-aligned inner test particle configuration, we also show results from another double averaged simulation(black) with initial conditions close to the original simulation(blue). These results are not in agreement with the original simulations which highlights the importance of chaos when the eccentricity is high.
\label{fig:emaxecc}}
\end{figure}

We use a numerical approach to study the maximum eccentricity excitation of initially nearly co-planar test particles under the influence of an eccentric perturber. In Figure \ref{fig:emaxecc} we show the maximum eccentricity as a function of semi-major axis ratio($\alpha$) for test particles initially aligned (upper panel) and anti-aligned (lower panel) with respect to an eccentric perturber. The perturber (with $m_2/m = 3 \times 10^{-4}$) is on an orbit with an  eccentricity of 0.6, and test particles start on an orbit with $e=0.4$ and $i=5^\circ$. 

We plot the maximum eccentricity for both inner (left panels) and outer (right panels) test particle configurations. In addition, we consider initially aligned configurations (top panels), where the test particle pericenter starts aligned with that of the perturber, as well as the anti-aligned configurations (bottom panels). The maximum eccentricity increases as the system becomes less hierarchical for both inner and outer test particles. Except for when the maximum eccentricity is close to 1 or when the test particles are close to mean-motion resonances, double averaged simulations agree with N-body simulations across all values of $\alpha$ shown here. We show maximum eccentricity as calculated using octupole hamiltonian for inner \citep{lithwick_eccentric_2011} and outer test particle configurations \citep{naoz_eccentric_2017} using green crosses. We can see that the octupole level hamiltonian provides a good approximation only for $\alpha<0.25$ and $\alpha>2$ in the initially aligned configuration and for $\alpha<0.1$ and $\alpha>8$ for the initially anti-aligned configuration.

It can be seen that when $e_{max}$ is close to 1, the double averaged results(blue) are not in agreement with N-body simulations(red) due to chaos. To illustrate the effects of chaos, we include results from another double averaged simulation with initial conditions close to the original simulation (black). We chose the new initial conditions by adding $10^{-4}$ to the original initial conditions in the phase space ($\{J,\omega,J_z,\Omega\}$). It can be seen that when the system becomes mildly hierarchical ($\alpha \gtrsim0.18$) the system becomes chaotic which leads to the discrepancy  between the secular and the N-body results. 

\begin{figure*}
\centering
\includegraphics[width=1.0\linewidth]{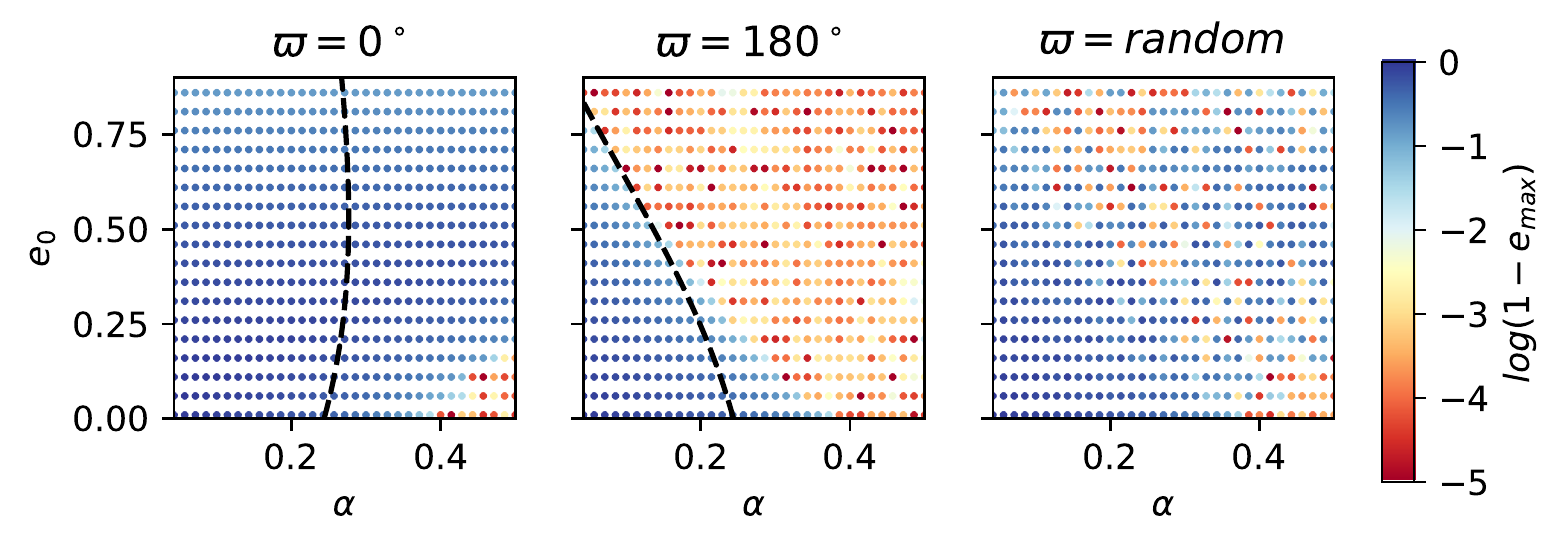} \\
\includegraphics[width=1.0\linewidth]{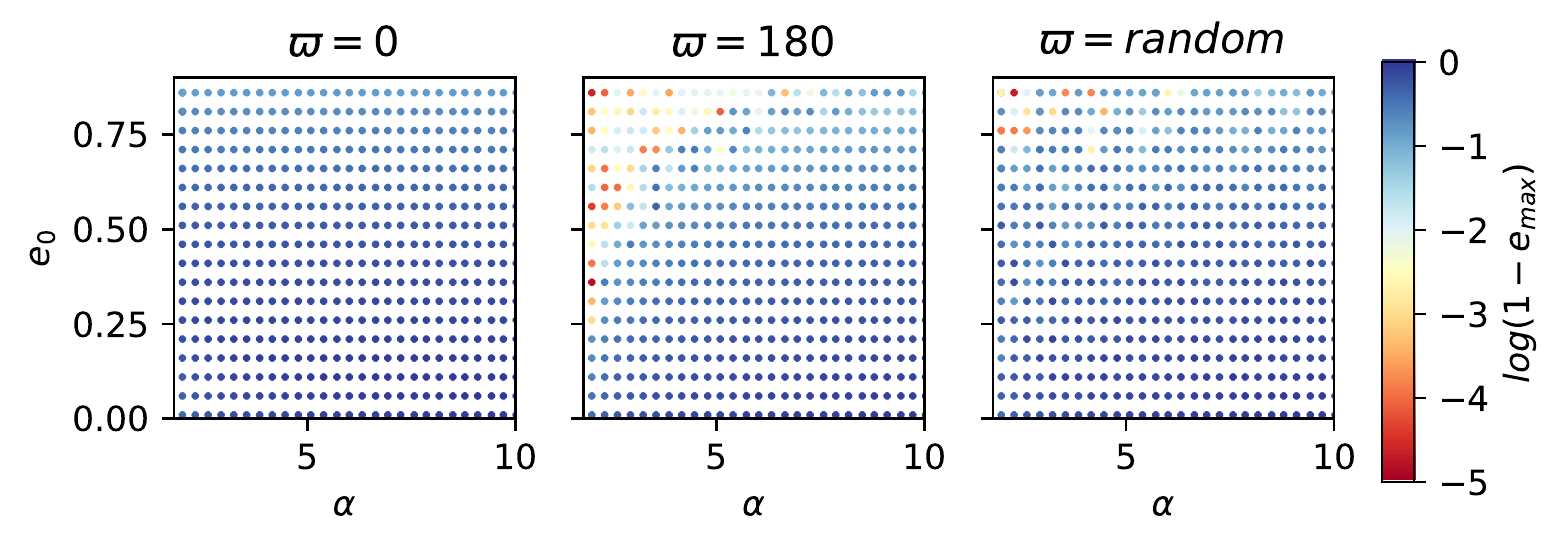}
\caption{Maximum eccentricity excitation for nearly coplanar, mildly hierarchical triples. x axis represents the ratio of semi-major axes and y-axis represents the initial eccentricity of the test particle. The dashed lines in the upper panels illustrate the orbit-flip criteria from \citep{li_eccentricity_2014}. Test particles start at an inclination of $5^\circ$. Perturber is on an orbit with $e=0.6,i=0,\omega=0$ and $\Omega=0$. Color represents the maximum eccentricity reached in 10 Kozai timescales.
The upper panels have an \emph{inner} test particle, the lower panels have an \emph{outer} test particle. 
\label{fig:eichare}}
\end{figure*}

\begin{figure*}
\centering
\includegraphics[width=1.0\linewidth]{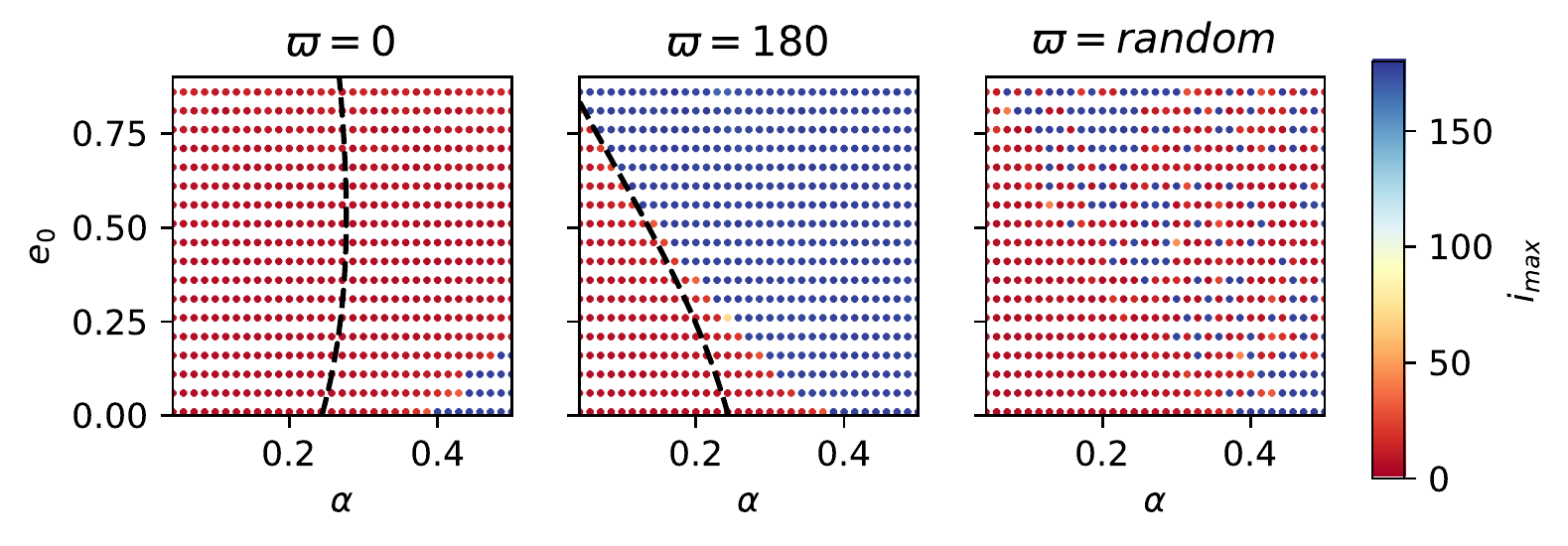} \\
\includegraphics[width=1.0\linewidth]{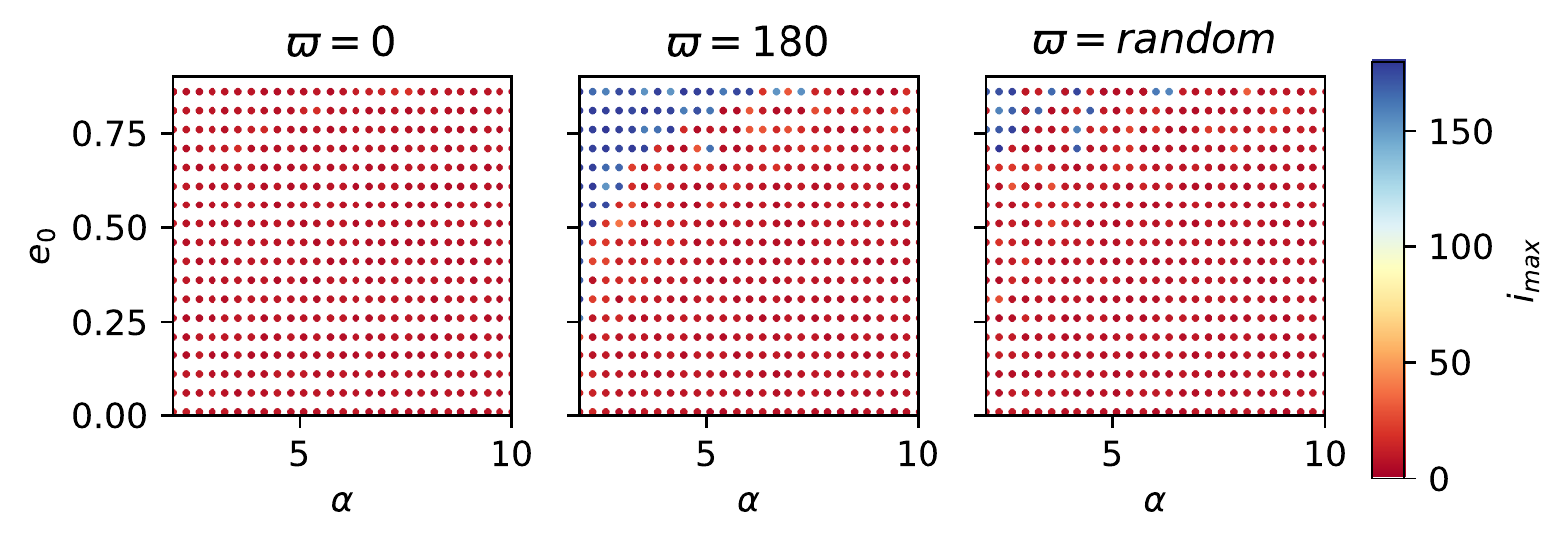} 
\caption{Maximum inclination excitation for nearly coplanar, mildly hierarchical triples. x axis represents the ratio of semi-major axes and y-axis represents the initial eccentricity of the test particle. Test particles start at an inclination of $5^\circ$. Perturber is on an orbit with $e=0.6,i=0,\omega=0$and $\Omega=0$. Color represents the maximum inclination reached in 10 Kozai timescales. The upper panels have an \emph{inner} test particle, the lower panels have an \emph{outer} test particle.\label{fig:eichari}}
\end{figure*}

To investigate the eccentricity and inclination variations for different initial eccentricities, we performed an ensemble of secular simulations. We choose the following initial conditions for the test particles: $\alpha \in [0.1,0.5]$ for inner test particle configurations and $\alpha \in [2,10]$ for outer test particle configurations, $e_0 \in \{0,0.9\}, i_0=5^\circ$ along with three different initial orientations $\varpi_0 =\{0^\circ,180^\circ,\text{random}\}$ where $\varpi$ is longitude of pericenter of the test particle and $e_0$ is the initial eccentricity of the test particle. For the perturber, we choose $e_2=0.6, \omega_2=0, \Omega_2=0$ and $i_2=0$. We evolve the inner-test particle systems for a maximum of 10$t_k$ and outer test-particle systems for 10$t_p$ and record the maximum eccentricity and inclination reached in each run. $t_k$ is the Kozai timescale (see Equation \ref{eq:tk}), and $t_p$ is the precession time scale given by:
\begin{eqnarray}
	t_p&=&\frac{(m+m_{pert})^2}{mm_{pert}} \frac{a_1^2}{a_{pert}^2} P \label{eq:tp},
\end{eqnarray} 
where $m$ is the mass of the central object, $m_{pert}$ is the mass of the perturber, $a_1$ is the semi major axis of the test particle, $a_{pert}$ is the semi-major axis of the perturber and $P$ is orbital period of the test particle\citep{murray_solar_1999}.
   
In Figure \ref{fig:eichare}, we show the maximum eccentricity in the plane of the initial eccentricity($e_0$) and semi-major axis ratio($\alpha$). The upper panels study the case of an inner test particle. The left panel starts with the aligned configuration, and the middle panel starts with the anti-aligned one. Similar to the hierarchical limit, starting from a nearly co-planar configuration, it is generally easier to excite the eccentricities as compared to initially aligned ones. In the hierarchical limit at the octupole order, an analytical expression can be derived to predict the flip of an initially nearly co-planar inner orbit \citep{li_eccentricity_2014}, and we include the criterion using dashed lines in the figures. It shows that the criterion can still be valid up to $\alpha \sim 0.1$ for the anti-aligned configuration.  

 The criterion does not predict the flips when the system is initially aligned since the orbit cannot be flipped starting in the aligned configuration in the hierarchical limit ($\alpha \lesssim 0.1$) at the octupole order \citep{li_eccentricity_2014}. The right panel corresponds to randomly aligned initial configurations ($\varpi=random$), and it shows the importance of the initial orientation. Specifically, the eccentricity can remain low for specific values of $\varpi$ even at high values of $\alpha$. This can be attributed to the resonant regions which can be seen in the surface of sections for high $\alpha$ configurations. 

The lower panel corresponds to the outer test particle configurations. In contrast to the inner test particle cases, starting from a nearly co-planar configuration, the eccentricity is rarely excited close to unity, which would lead to the orbits to cross for the outer test particles. This is consistent with earlier results in literature which show that up to quadrupole order of the disturbing function, the eccentricity of the outer test particle is constant\citep{naoz_eccentric_2017,vinson_secular_2018}.  Similar to the inner test particles , starting from a nearly co-planar configuration, the eccentricity is more likely to be excited when the orbits are anti-aligned, and when the system is less hierarchical. Only in anti-aligned configurations can the near circular orbit excite to high eccentricities, even in the mildly hierarchical configurations. 

It is also apparent that large inclination variations are accompanied by large eccentricity excitations. To illustrate the variations in inclination, we color code the maximum inclination as a function of the initial eccentricity ($e_0$) and semi-major axes ratio ($\alpha$) in Figure \ref{fig:eichari}. The flip criterion in the octupole limit by \citet{li_eccentricity_2014} is included as black dashed lines, and it is consistent with the double averaged results for $\alpha<0.1$ and $\varpi=180$. At higher values of $\alpha$, the orbits can flip with lower initial eccentricities as compared to the flip criteria. This is consistent with the results shown in Figure \ref{fig:emaxecc}. In particular, starting from a near co-planar configuration, a near circular inner test particle orbit could flip due to the perturbation of an eccentric outer object starting anti-aligned with the test particle with $\alpha \sim 0.4$. 

For outer test particles (lower panel), starting from a near co-planar configuration, orbit flips are rare unless the orbit starts mildly hierarchical ($\alpha \lesssim 10$) and in an anti-aligned configuration. This is consistent with \cite{naoz_eccentric_2017} who find that in hierachical systems starting from a nearly co-planar configuration, inclination cannot be excited to flip the orbit of the outer particle. Eccentricities are excited close to unity during the flips, and frequently causes the cross of orbits. Similar to the eccentricity excitation, the flips are only seen when the orbits are anti-aligned. 


\section{Applications to outer solar system} 
\label{section:outersol}

Mildly hierarchical triples are stable when the perturbing mass is much less massive than the central object, such as our own Solar System's extreme trans-Neptunian objects (eTNOs, with large semi-major axes and high eccentricity) under the perturbation of a possible undetected planet residing far from the Sun (Planet-9). In this section, we apply our understanding of the dynamics of the mildly hierarchical triples to the Solar System, and constrain the properties of the possible undetected Planet-9. 

Planet-9 is proposed to explain mysterious features in the orbital distribution of extreme trans-neptunian object (objects that lie outside the orbit of Neptune with large semi-major axes $\gtrsim 150$ AU) \citep{sheppard_new_2016, batygin_evidence_2016}. In particular, they show clustering in the longitude of the ascending node, the argument of pericenter and their orbital planes. Many studies have shown that the alignment of the orbits is not due to selection biases \citep{deLaFuenteMarcos14, Gomes15, batygin_evidence_2016}, although \citet{shankman_consequences_2017} demonstrated that the “Outer Solar System Origins Survey” (OSSOS; \citealt{bannister_ossos._2016}) contains nonintuitive biases for the detection of TNOs that lead to apparent clustering of orbital angles in their data, and the angular elements of the distant TNOs are consistent with uniform distribution \citep{bannister_ossos._2018}. Ongoing observational TNO surveys will provide a better understanding of the architecture of the outer solar system and the details of the TNO clustering, and it's important to obtain a better understanding on the dynamical origin of possible clusterings of eTNOs. 

The proposed Planet-9 is on an eccentric orbit with a large semi-major axis which can perturb the orbits of the eTNOs and explain the clustering of the eTNOs. The underlying dynamics of the interactions between the eTNOs and Planet-9 is rich, and it is found that secular dynamics plays an important role leading to the clustering of the eTNO orbits \citep{hadden_chaotic_2018, saillenfest_long-term_2017, li_secular_2018}. To illustrate the dominance of the secular resonances, we compare our secular results with the those we got from N-body simulations below. 

\begin{figure}
\includegraphics[width=\linewidth]{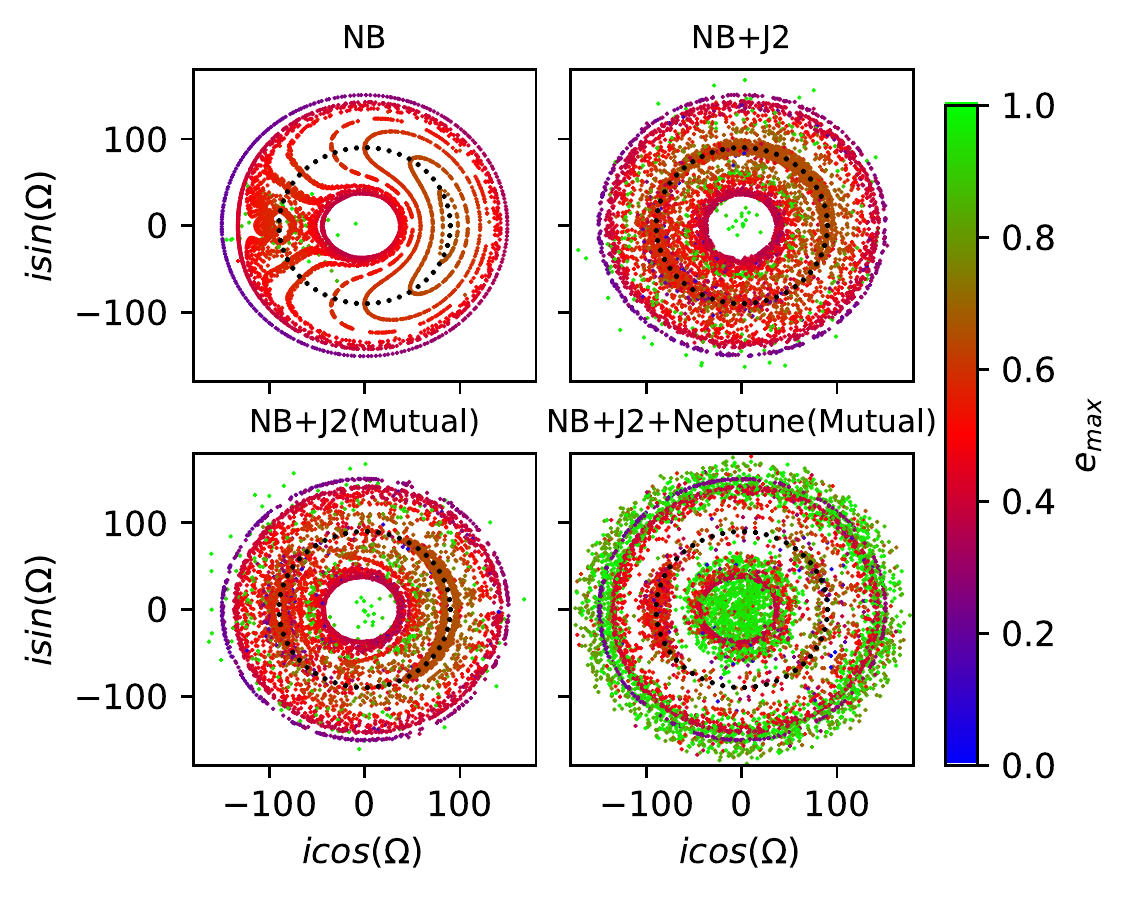}
\caption{Comparison of surfaces for test particles with $a=250$ AU and perturber with $a_9=500$AU, $e_9=0.2$ and $i_9=0$.  The dashed circle in black represents an inclination of 90$^\circ$. {\bf Top left panel}: Projections in $i\cos{\Omega}-i\sin{\Omega}$ using N-body simulation, {\bf Top Right panel}: Projection evaluated using N-body code with a $J_2$ potential to account for interactions with the giant planets. {\bf Bottom Left panel}: Projection with the $J_2$ potential in mutual coordinates. In mutual coordinates, the location of resonance is restored to $\Omega=0 \text{and }\pi$. {\bf Bottom Right panel}: Projection evaluated in mutual coordinates using N-body code with Neptune as an active object and a $J_2$ potential for other planets of the solar system. }
\label{fig:secnbd}
\end{figure}

In Figure \ref{fig:secnbd}, we show the projections for particles orbiting around the Sun at $a=250$AU, under the perturbation of a Planet-9 of mass $m_9 = 10 M_{\oplus}$, with semi-major axis $a_9=500$AU and eccentricity of $e_9=0.2$. To illustrate the inclination variations and the orbital plane clustering, we show the projections in the plane of $i\sin(\Omega)$ and $i\cos(\Omega)$. The projections are obtained in the same way as the surface of sections in Section \ref{section:soc}. We choose the initial condition of the test particle based the results of the surface of section, so that they start with the same energy. The color represents the eccentricity of the particles. The top-left panel shows the results from N-body simulations, which do not include the effects of the giant planets. Without the effects of the giant planets, the results of the three-body secular results largely resembles that of the N-body simulations (compare with the last panel for $\alpha=0.5$ in Figure \ref{fig:i2}), except that parts of the libration region around $\Omega \sim 180^\circ$ become chaotic. This is due to orbital crossing of eTNOs with Planet-9, which is more often when the orbits' pericenters are anti-aligned with repect to each other and lead to close scattering of eTNOs. The close scattering of eTNOs by Planet-9 cannot be reproduced by the secular results, and it often leads to chaotic behaviors.

In the top-right panel, we show N-body results which include the effects of the giant planets as a $J_2$ potential. The $J_2$ potential causes precession in $\Omega$, which leads to more frequent orbital crossing. This suppresses the dominance of resonances in $\Omega$ and leads to larger chaotic regions. However, eccentricity excitation near $\Omega \sim 0^\circ$ due to the secular resonances can still be found in the N-body results, and this can produce the clustering of the eTNO orbits with high eccentricity. In addition, with the $J_2$ potential, the orbit of the Planet Nine is no longer constant. It undergoes precession and this makes a slight shift in the resonant region at $\Omega=180^\circ$. 

In the secular results, the orbital elements are obtained relative to the orbit of the outer planet, however the orbital elements are relative to the ecliptic plane in the N-body results. While the orbital elements of the pertuber in non-hierarchical triples are constant, Planet-9's orbital elements are not. Neptune and other inner planets exert torques which causes Planet-9's orbit to precess. Thus, to better compare our secular results for mildly hierarchical triples with N-body simulations of eTNOs with Planet-9 and the $J_2$ potential, we use mutual coordinates, which is set up as the following. We choose our reference plane and reference direction such that inclination($i_9$), argument of pericenter($\omega_9$) and longitude of ascending node($\Omega_9$) of the Planet-9 are set to 0, following the example of \citet{gronchi_generalized_2002}. The results based on the mutual coordinates are shown in in the bottom-left panel, and we can see that the resonance at $\Omega=0$ is restored in mutual coordinates. The precession due to the $J_2$ potential allows the orbit of the perturber and that of the test particles to cross more frequently. This causes the system to become significantly more chaotic as shown in Figure \ref{fig:secnbd}.

Finally, we include a point mass Neptune in our N-body simulation (as an active body instead of a $J_2$ term) and show the results in the bottom right panel of Figure \ref{fig:secnbd}. Including a point mass Neptune, the secular dynamical features get further suppressed. However, we could still see some clustering of the orbit in $\Omega$ with moderate eccentricity at high inclination around $\sim 180^\circ$, due to the secular resonances.

We next ran an ensemble of N-body simulations and constrain the properties of Planet-9. Multiple studies have attempted to constrain the orbital parameters of a potential Planet-9 using N-body simulations. In their original paper \cite{batygin_evidence_2016} reported that Planet-9 could be a 10$M_\oplus$($m_9$) planet on an orbit with a semi-major axis($a_9$) of 700 AU,  an eccentricity of 0.6($e_9$) and an inclination($i_9$) of $30^\circ$. These values were updated in a recent paper, where \cite{batygin_planet_2019} find that $m_9\sim (5,10)$M$\oplus$, $a_9 \sim (400,800)$AU, $e_9 \sim (0.2,0.5)$ and $i_9 \sim (15-25)^\circ$. The previous studies have focused on configurations of Planet-9 with inclination less that $35^\circ$ , but it is possible that Planet-9 lies largely out of the plane. Thus, we sample the inclination of Planet-9 over a wide range of inclinations here. 

In Table \ref{tab:p9oe} we list the configurations of Planet-9 we use for the N-body simulations. We use 1000 test particles to model the Kuiper belt. Initial conditions for these particles are listed in Table \ref{tab:tpoe}. We also include Neptune as a point particle in our simulations. Other giant planets in the solar system are modeled using a $J_2$ potential. We use the hybrid symplectic integration method in \texttt{MERCURY} package to do our simulations. Following \cite{batygin_planet_2019}, we use a time-step of $10\%$ of the orbital period of the Neptune.
\begin{table}
\centering
\begin{tabular}{ |c| } 
 \hline
 $a_9 \in \{300,700,1000,1400\} \text{AU}$\\ 
 \hline
 $e_9 \in \{0.2,0.4,0.6,0.8\}$\\ 
 \hline
 $m_9 \in \{5,10,20,30\} M_\oplus$\\ 
 \hline
 $i_9 \in \{0,30^\circ,60^\circ,90^\circ,120^\circ,150^\circ,180^\circ\}$ \\
 \hline
 $\omega_9=0,\Omega_9=0$ \\
 \hline
\end{tabular}
\caption{\label{tab:p9oe}Planet-9 orbital elements. We note that some of the parameters above could be ruled out by observational constraints, but we included them here in order to analyze the dynamics in a broad range of parameter space.}
\end{table}

\begin{table}
\centering
\begin{tabular}{ |c|c|c| } 
 \hline
 Orbital element & Range & Distribution \\
 \hline
	$a_t$ & $[100,800] \text{AU}$ & Uniform\\ 
 \hline
	$q_t$ & $[30,100] \text{AU}$ & Uniform\\ 
 \hline
 $\omega_t$ & $[0^\circ,360^\circ]$ & Uniform\\ 
 \hline
 $\Omega_t$ & $[0^\circ,360^\circ]$ & Uniform\\
 \hline
 $i_t $ &  $[0^\circ,180^\circ]$ & Half-Normal with $\sigma=15^\circ$ \\
 \hline
\end{tabular}
\caption{\label{tab:tpoe}Test particle orbital elements.}
\end{table}

We look for configurations of Planet-9 which would produce the clustering of eTNO orbits. Currently, eight out of nine metastable eTNOs have $\varpi \in [330,250]^\circ$ and inclination and longitude of ascending node with $<i>=7^\circ, <\Omega>=32^\circ$ and $\sigma_i=15 ^\circ$ \citep{batygin_planet_2019}. 

\begin{figure}[h]
\includegraphics[width=1.0\linewidth]{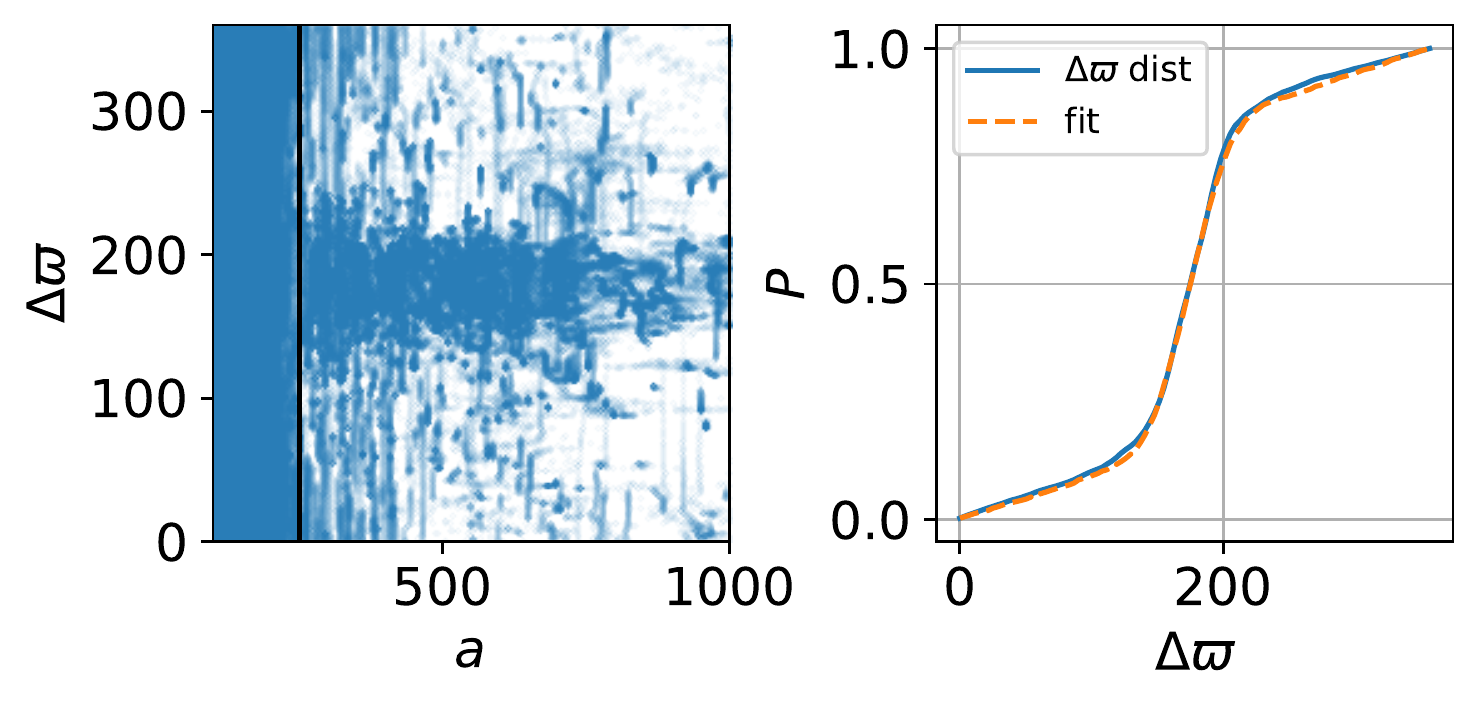}
\caption{Clustering in the longitude of pericenter ($\Delta \varpi$) ({\bf left}) and fitting of the $\Delta \varpi$ distribution ({\bf right}). Test-particles after 2 billion years of evolution are included with time step of 1Myr. Black line shows the critical semi-major axis. Test particles with $a>a_c$ show clustering in $\Delta \varpi$. }
\label{fig:plane_clust}
\end{figure}

We first calculate the critical semi-major axis $a_c$, beyond which we expect to see clustering. We do this by dividing the simulation data into bins in semi-major axis and mark the bin where the distribution of $\varpi$ stops being uniform. In our analysis we look for clustering in the longitude of pericenter by fitting the distribution of $\Delta \varpi (=\varpi-\varpi_9)$ for $a>a_c$ using the following probability density function:
\begin{equation}
	P(x)= (\kappa) \text{uniform}(0,360) + (1-\kappa) \text{normal}(x,\mu,\sigma),
\end{equation}
where $\kappa$ is a measure of the strength of the clustering, where a larger value of $\kappa$ corresponds to less clustering. $\mu$ and $\sigma$ are the mean and the standard deviation of the normal distribution, which denote the location and the strength of the cluster. We calculate the values of $\kappa,\mu$ and $\sigma$ for all the simulations we do, and we use the Kolmogorov-Smirnov test to check the goodness of fit. We select the simulations which pass the test in the following analysis.

In Figure \ref{fig:plane_clust} we show an example. In the left panel we plot $\Delta \varpi$ as a function of the semi-major axis ratio for test particles. In this simulation, Planet-9 has a semi-major axis of $700$ AU with an eccentricity of 0.6, inclination of $0^\circ$ and mass of 10$M_\oplus$. We show the critical semi-major axis ratio in black at 260 AU for this simulation. The points are sampled from the trajectories of the test-particles. Points after 2 billion years of evolution are chosen at an interval of 1 million years. On the right we show the show the distribution of $\Delta \varpi$ for $a > a_c$ (where $a_c=260$ AU), as well as the fit. For this fit we find $\mu=177.7^\circ, \kappa=0.321 $ and $\sigma=23.70^\circ$.

We use association rule analysis to determine which configurations of Planet-9 leads to clustering of the test particles (lower $\kappa$). 
More specifically, association rule analysis is a machine learning method used to find relationships between variables in a data set. In the context of this study, each N-body simulation corresponds to a transaction which comprises of the orbital parameters of Planet-9 and the value of the fitting parameter $\kappa$, calculated using the method described above. To use the association rule analysis we discretize the parameter $\kappa$ as the following:
\begin{eqnarray}
	0   < 1-\kappa < 0.3 &\rightarrow& \text{low cluster}, \nonumber \\
	0.3 < 1-\kappa < 0.6 &\rightarrow& \text{medium cluster}, \nonumber \\
	0.6 < 1-\kappa < 1 &\rightarrow& \text{high cluster}. \nonumber
\end{eqnarray}
For example, for the simulation with Planet-9 orbital elements $\{a_9=300 \text{AU}, e_9=0.6, i_9 =0, m_9=5\}$ we got $\kappa=0.55$. This translates to a transaction $t=\{``a_9=300\text{AU}", ``e_9=0.6", ``i_9=0", ``m_9=5"$ ,``medium cluster"\}.

Association rules are denoted by the notation: $A \rightarrow B$, where $A$ and $B$ are sets of items. Two quantities are used to specify which rules are interesting: support ($s$) and confidence ($c$). Support indicates how frequently a set appears in a data set, and confidence indicates how often the rule is valid. For a given rule $A\rightarrow B$, they are:
\begin{equation}
	s=\sigma(A\cup B)/|D|, c=\sigma(A \cup B)/\sigma(A),
\end{equation}
where $\sigma(A)$ is the count of the set $A$ (i.e. number of times set $A$ appears in the data set) and $|D|$ is the number of transactions in the data set. In this case, $|D|$ is the number of simulations which satisfy the fitting criteria for $\kappa$. Rules with higher support are more widely applicable. Confidence measures the reliability of a rule, and rules with higher confidence are more reliable. 

Using rules with $c=1.0$, we find the parameter space that always leads to orbital clustering. For instance, we find:
\begin{eqnarray}
	\{a_9=1000\text{AU},e_9=0.8,i_9=90^\circ\} &\rightarrow& {\text{\{high cluster\}}},\nonumber \\ 
	&&(c=1.0,s=0.011) \nonumber.
\end{eqnarray}
This rule tells us that irrespective of what the mass of Planet-9 is, configurations with $a_9=1000\text{AU},e_9=0.8$ and $i_9=90^\circ$ lead to high clustering in $\varpi$.

Other rules with high confidences are listed below:
\begin{eqnarray}
	\{a_9=700,i_9=0,m_9=5\} &\rightarrow& \text{\{high cluster\}}\nonumber \\
	 &&(c=1.0,s=0.0073), \nonumber\\
	\{a_9=1400,e_9=0.8,i_9=90^\circ\} &\rightarrow& \text{\{high cluster\}}\nonumber \\
	&&(c=1.0,s=0.0073), \nonumber\\
	\{e_9=0.2\} &\rightarrow& \text{\{low cluster\}} \nonumber \\
	&&(c=0.86,s=0.21), \nonumber\\
	\{m_9=30\} &\rightarrow& \text{\{low cluster\}} \nonumber \\
	&&(c=0.71,s=0.16). \nonumber
\end{eqnarray}
The specific values of the support indicate the size of the parameter space that lead to the clustering based on the rules. Thus, rules with fewer Planet-9 parameters (e.g., the third and the fourth rule above) tend to have higher support. While rules with higher support can be applied widely, they tend to have lower confidence, which limits their validity. It should be noted that confidence and support are calculated only for simulations in which $\Delta \varpi$ distribution was a good fit for the test distribution we defined above. Of the 448 simulations we ran, only 276 fit this criteria. 

It is intriguing that the high inclination Planet-9 with large semi-major axis ($a_9 \sim $1000-1400 AU) on an eccentric orbit($e_9 \sim 0.6-0.8 $) could lead to strong clustering in $\varpi$. This is due to the long secular oscillation timescale of $\varpi$. During the 4 Gyr simulation, $\varpi$ starts to converge to $\sim 170^\circ$ and leads to the clustering.

\begin{figure}
\includegraphics[width=1.0\linewidth]{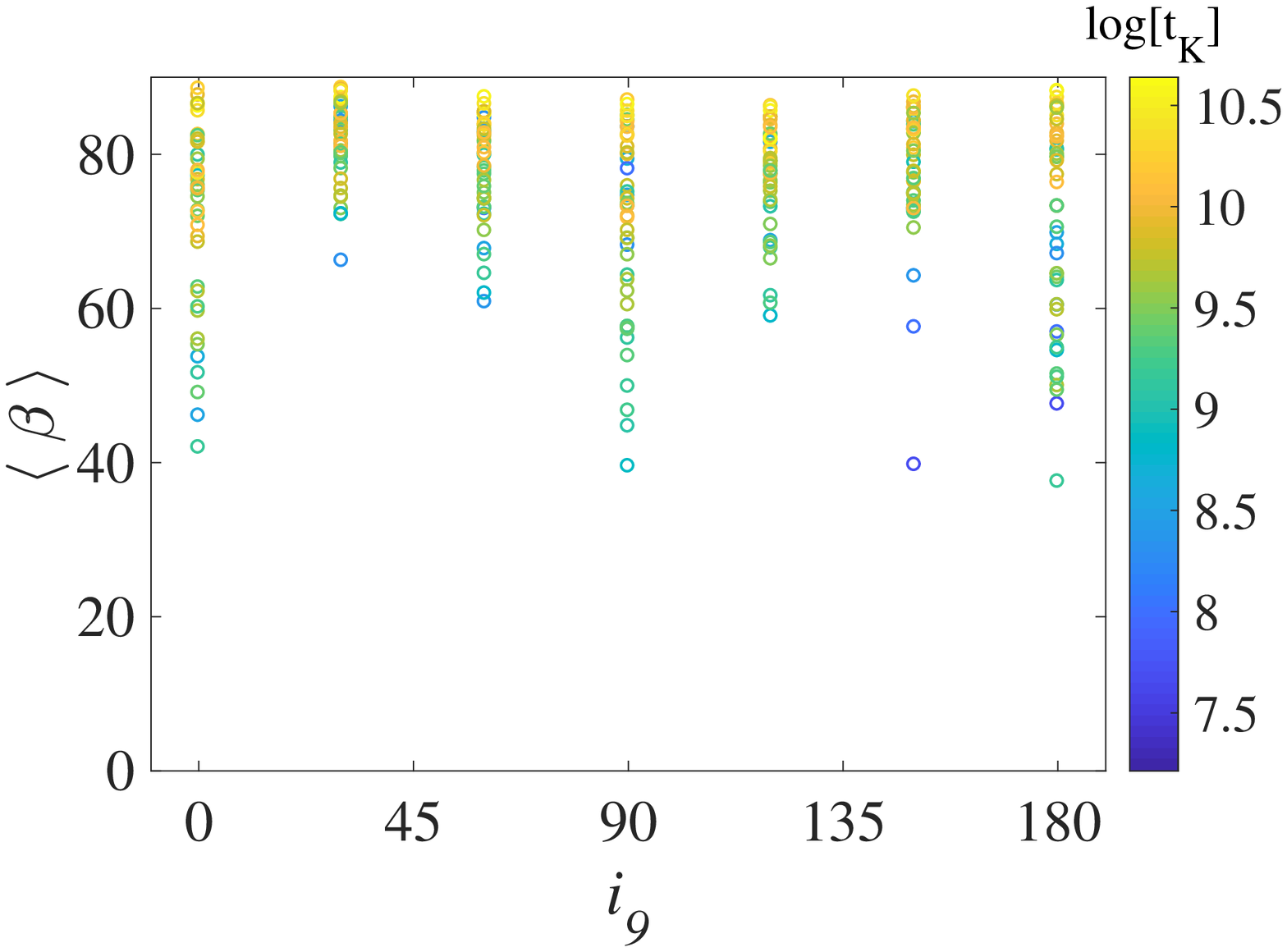}\\
\includegraphics[width=1.0\linewidth]{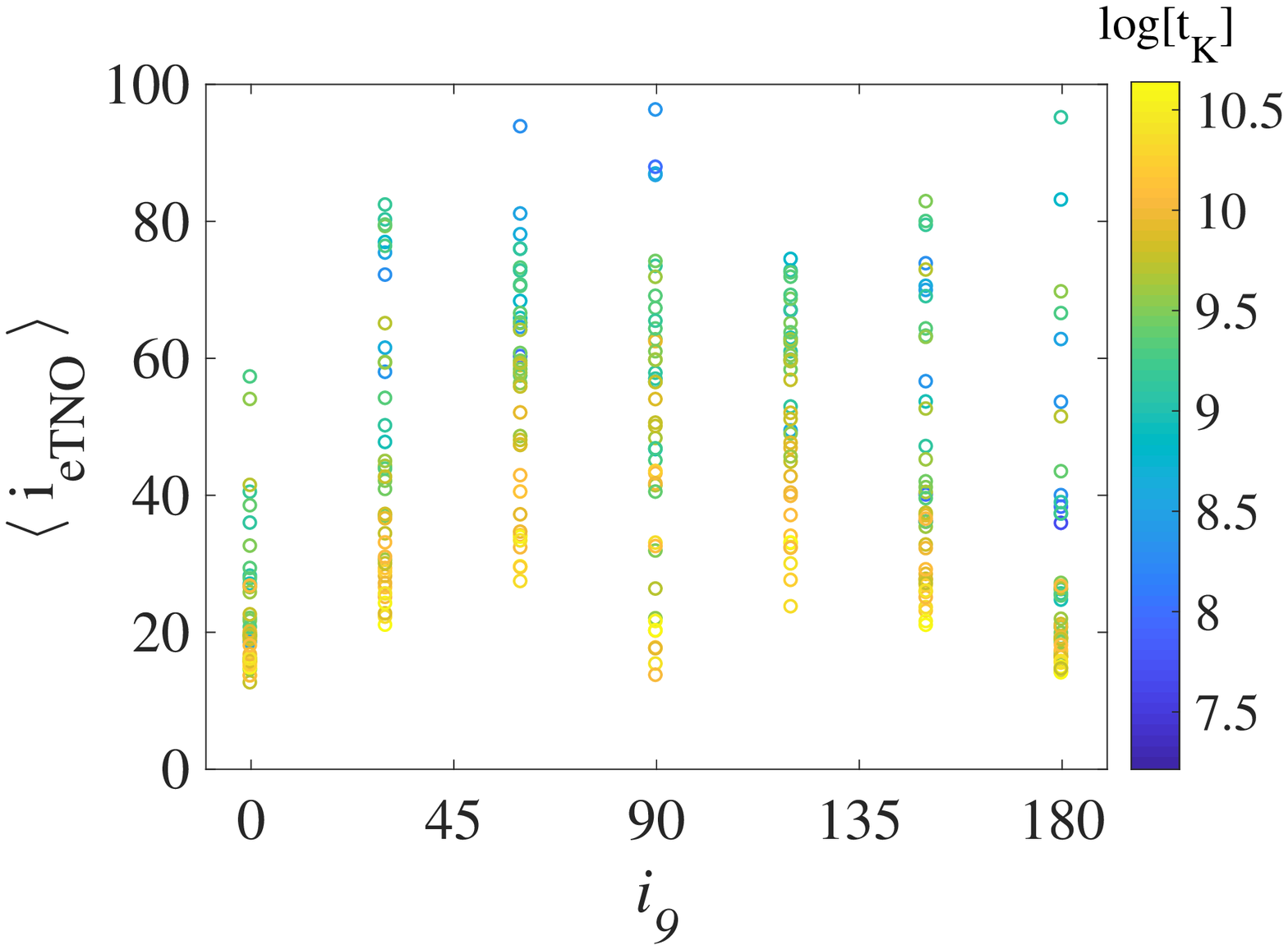}
\caption{Clustering and average inclination of eTNOs ($30<r_p<100$AU, $a<250$AU) as a function of Planet-9's inclination. {\bf Upper panel}: Mean angular spread of pericenter directions $\langle \beta \rangle$ (eqn \ref{eq:beta}) as a function of Planet-9 inclination. {\bf Lower panel}: Mean eTNO inclination as a function of Planet-9 inclination. Color represents the Kozai timescale for a particle with $a=250$AU. It shows that near coplanar, polar ($\sim 90^\circ$) and retrograde Planet-9 with short $t_K\sim1$Gyr could lead to the clustering of the pericenter directions, and both near coplanar and counter-orbiting Planet-9  with short $t_K\sim1$Gyr could produce low average eTNO inclination.}
\label{fig:planeclust}
\end{figure}


With high inclinations, the cluster in $\varpi$ may not correspond to clustering in the orbital directions geometrically. To study the geometrical clustering, we find the spread in the pericenter directions directly, following the approach by \citealt{millholland_obliquity-driven_2019}. We select particles with pericenter distance between $30$ and $100$AU and semi-major axis above $250$AU for this analysis, since the scattering with the giant planets becomes important when $r_p<30$AU, and it is challenging to detect the eTNOs with $r_p>100$AU. Below $a=250$AU, the clustering is weak due to the fast precession in the $J_2$ potential of the giant planets. 

We first estimate the average pericenter direction as the following:
\begin{equation}
	\langle\hat{\bf{e}}\rangle = \frac{\sum_i^N \hat{\bf{e}}_i}{|\sum_i^N \hat{\bf{e}}_i|},
\end{equation}
where $\hat{\bf{e}}_i$ is the eccentricity unit vector of $i^{th}$ test particle, and $N$ is the number of test particles which survived 4 billion years of evolution residing in the selected parameter region ($30<r_p<100$AU, $a>250$AU). Then, we calculate $\beta_{i}$, which is the angle between the pericenter orientation of the $i^{th}$ test particle and the average pericenter direction. As a measure of the pericenter clustering, we calculate the average values of $\beta_{i}$ denoted by $\langle\beta\rangle$:
\begin{equation}
	\langle \beta \rangle = \frac{\sum_i^N \arccos(\hat{\bf{e}}_i \cdot \langle \hat{\bf{e}} \rangle)}{N}.
	\label{eq:beta}
\end{equation}
Since the particles start in the same plane, where $\beta_i$ is uniformly distributed between $0^\circ$ and $180^\circ$, the initial $\langle \beta \rangle$ is  $\sim 90^\circ$. Using the same approach for the observed sample using data from the Minor Planet Center\footnote{https://www.minorplanetcenter.net/}, there are currently 19 eTNOs with $\langle \beta_{obs} \rangle = 60^\circ$, lying with low inclination $\langle i_{obs}\rangle = 18^\circ$.

In the upper panel of Figure \ref{fig:planeclust}, we plot $\langle\beta\rangle$ against inclination of Planet-9. To ensure the results to be statistically robust, we only select simulations with more than $50$ particles left in the selected region ($30<r_p<100$AU, $a>250$AU). Color denotes the Kozai timescale due to the perturbations induced by Planet-9 (Eqn. \ref{eq:tk}) for an eTNO with semi-major axis of $250$AU. It shows that near coplanar, polar and counter orbiting Planet-9 all could lead to the clustering in the pericenter directions, with short dynamical timecales $t_K \lesssim 1$Gyr. In addition, Planet-9 with shorter Kozai timescales could also lead to stronger clustering. The retrograde orbiting Planet-9 with the same dynamical timescales lead to the same results as that of the prograde cases. This is because the secular dynamics is independence on the direction of the motion of Planet-9. 

When Planet-9 is highly inclined and retrograde ($i_9 \sim 150^\circ$), the massive Planet-9 ($\sim 30$M$_\oplus$) with very short dynamical timescale $t_K\sim 10$Myr ($a_9 \sim 300$AU), could also lead to clustering of the pericenters. However, this feature is missing in the prograde scenario, since not enough particles are left in the selected region as the prograde Planet-9 is more disruptive due to the non-secular effects. Nevertheless, Planet-9 in this parameter space are unlikely since the high mass Planet-9 with small and eccentric orbits can mostly be ruled out observationally.

On the lower panel, we plot the average inclination of the selected test particles ($30<r_p<100$AU, $a>250$AU) versus the inclination of Planet-9. We find the the average inclination can further constrain the inclination of Planet-9. Although both the polar and the planar Planet-9 could produce the clustering, the average inclination produced by the polar Planet-9 is much higher. The current observation of the eTNOs with low average inclination orbits ($\sim 18^\circ$) favors the near co-planar and counter orbiting Planet-9. We note that the observational surveys are not complete for the high inclination region. The spread of eTNO inclinations due to a polar Planet-9 is large ($\sim 40^\circ$), and selecting only those below $40^\circ$ still shows strong clustering. Thus, current observations cannot yet rule out a polar Planet-9, and future observations on the high inclination eTNOs are needed to better constrain the inclination of Planet-9.


\section{Conclusions}
\label{section:conc}


In this paper we study the secular dynamics of mildly-hierarchical triple systems, and apply it to the dynamics of eTNOs in the outer Solar System. In contrast to the hierarchical limit, where one could use a perturbative approach to analyze the dynamics analytically by expanding in semi-major axes ratio, the dynamics can only be investigated numerically for mildly hierarchical systems. Thus, we developed code to numerically calculate the double averaged Hamiltonian, and to evolve the system to study its secular dynamics. The code is now publicly available (https://github.com/bhareeshg/gda3bd). 

In mildly hierarchical systems, the perturbation of a massive object can lead to oscillations that render the secular double averaged results unreliable. However, the oscillation amplitude is small and secular results still provide a good approximation when the perturber mass is low. To illustrate this, we calculate the maximum eccentricity induced by the secular interactions and compare that with the N-body simulation in Section \ref{section:Val}. We obtained an analytical expression for the critical mass, below which the secular results are reliable (Eqn. \ref{eq:mccri}) for systems with a circular outer perturber. When the semi-major axes ratio is small $\alpha \sim 0.1$, the critical mass is $m_2/m \sim 0.05$. With higher values of semi-major axes ratio ($\alpha \sim 0.5$), secular results hold for $m_2/m < 10^{-3.5}$. The critical mass decays with $\alpha^{-3/2}$. 

In addition, we make surface of section plots to study the dynamics of mildly-hierarchical triples and compare these with the hierarchical limit. We find that the secular dynamics have a weak dependence on the semi-major axis ratio ($\alpha$) if the perturber does not have a high eccentricity ($e_2 \lesssim 0.8$). For inner test particles, we find secular resonances at $\Omega=\{0,\pi,\pi/2,3\pi/2\}$ and $\omega=\{\pi,\pi/2,3\pi/2\}$, and for outer test particle we find secular resonances at $\Omega=\{\pi,\pi/2,3\pi/2\}$ and $\omega=\{\pi/2,3\pi/2\}$. This is similar to the hierarchical limit. The chaotic regions increase as the systems become less hierarchical. In contrast to the hierarchical limit, test particle orbits can flip from a near co-planar configuration starting with low eccentricities.

We then study the eccentricity and the inclination excitation of the test particles inside the mildly-hierarchical triples. For circular perturbers, we obtain semi-analytical results for the maximum eccentricity with an $8^{th}$ order expansion in the semi-major axes ratio, which agrees with the N-body results up to semi-major axes ratio of $\alpha \sim 0.3$ (section \ref{section:Val}). For an eccentric perturber, double averaged results and N-body results agree well with each other, while octupole results typically give higher $e_{max}$ when $\alpha\gtrsim0.1$. We then perform an ensemble of secular simulations, and find that the eccentricity is easily excited for initially anti-aligned systems and for less hierarchical systems. 

Finally, we apply our results to objects in the outer solar system. The secular dynamical features resemble that of the N-body results for the eTNOs under the perturbation of the undetected outer planet (Planet-9). Using surface of section, we find that the secular resonances can lead to clustering in eTNO orbits. Next, we perform an ensemble of N-body simulations to model the interactions between Planet-9 and eTNOs. In addition to a low inclination Planet-9 as identified in the literature \citep[e.g.,][]{batygin_planet_2019}, we find that a polar ($i_9\sim 90^\circ$) and a counter orbiting ($i_9\sim 180^\circ$) Planet-9 with Kozai timescales of $\lesssim 1$Gyr (for an eTNO with $a=250$) could also produce strong clustering. The high inclination or counter orbiting Planet-9 could post challenges on the formation of such a highly inclined wide orbit planet, which may indicate that Planet-9 could be a captured planet from another star \citep{Li_P9origin_16}. The near-coplanar and the counter orbiting Planet-9 could lead to low inclination eTNOs (average inclination of $\sim 20^\circ$), while the polar one leads to higher inclination  $\sim 60^\circ$ eTNOs with a large spread $\sigma_i \sim 40^\circ$. Future observations of the eTNOs with high inclinations could better constrain the Planet-9 inclination.
\section*{Acknowledgement}
The authors thank Evgeni Grishin, Melaine Saillenfest and the referee for helpful comments, which greatly improved the quality of the paper. This work was supported by NASA grant 80NSSC20K0641 and 80NSSC20K0522.

\appendix

\section{\textbf{A1. Numerical accuracy and integration methods}}

\begin{figure}
    \centering
    \begin{minipage}{0.45\textwidth}
        \centering
        \includegraphics[width=0.9\textwidth]{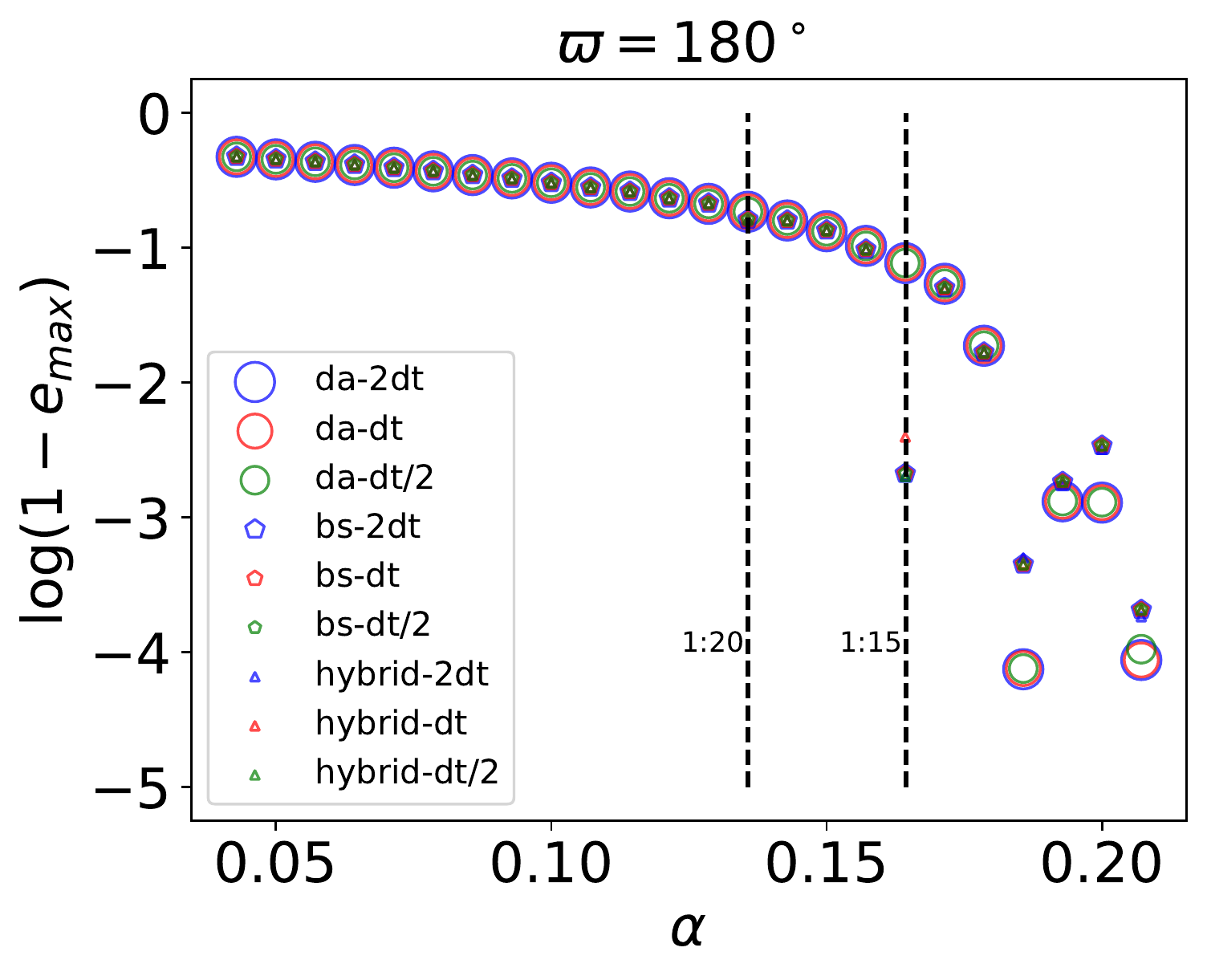} 
    \end{minipage}\hfill
    \begin{minipage}{0.45\textwidth}
        \centering
        \includegraphics[width=0.9\textwidth]{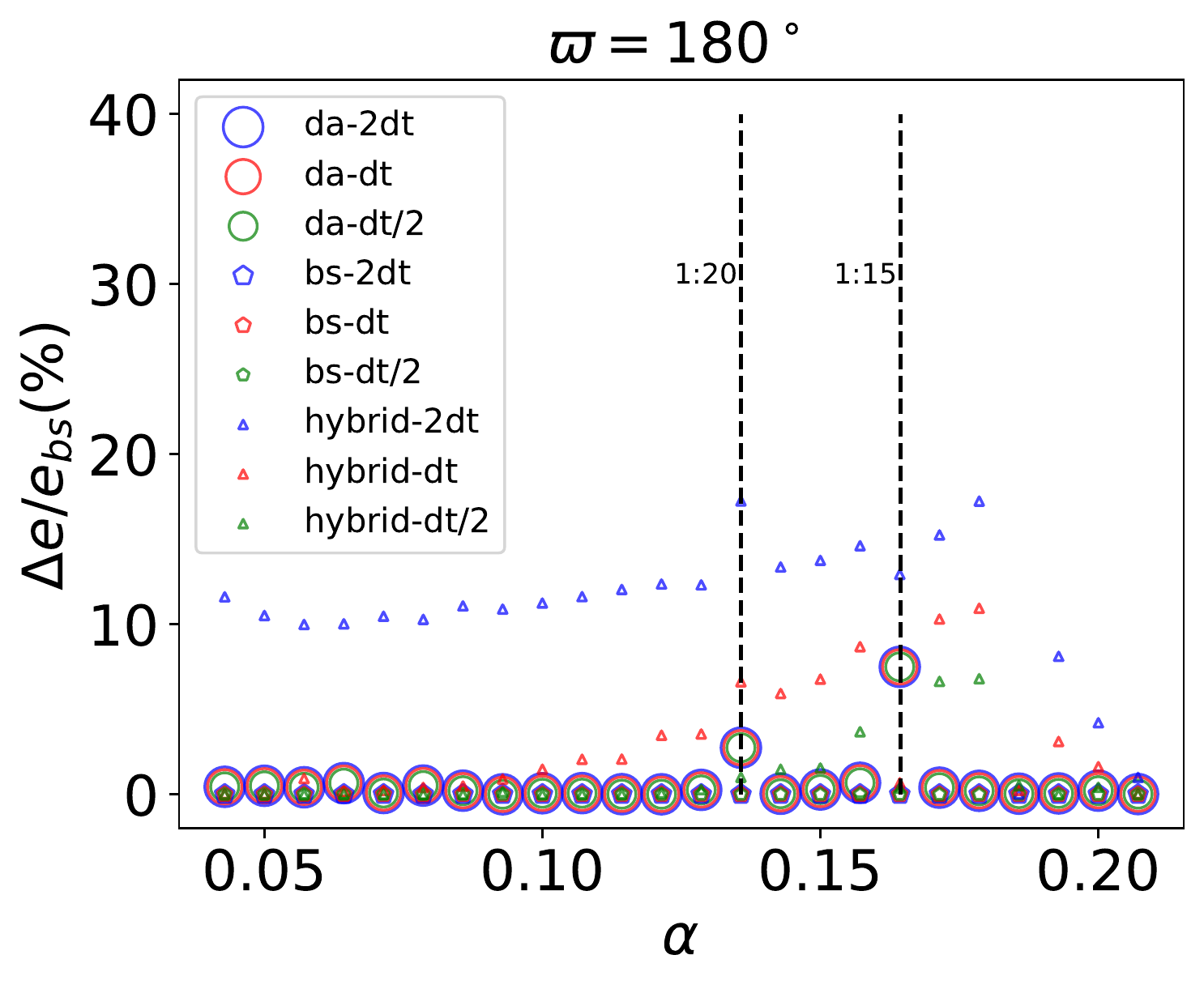} 
    \end{minipage}
    \caption{Comparison of various integration schemes and timesteps. {\bf Left Panel}: We plot the maximum eccentricity as a function of semi-major axis ratio for different integration schemes and timesteps. We show results for Bulirsch-Stoer integrator(bs) and hybrid integrator(hybrid) with timesteps of $2.5\%$, $5.0\%$ and $10\%$ of the period of the innermost orbit. {\bf Right Panel}: We show the relative difference in maximum eccentricity. Hybrid integration scheme does not agree very well at high eccentricity. Also, double averaged results(da) agree with N-body results when the system is not near mean motion resonances(black dashed lines). Initial conditions are same as those used in Figure \ref{fig:emaxecc}.}
     \label{fig:nerr}
\end{figure}

To ensure consistency in our N-body results, here we compare various integration schemes and choices of timesteps. In Figure \ref{fig:nerr} on the left panel, we show the maximum eccentricity (reached in $3 t_K$) as calculated using the double averaged simulation(da) and N-body simulations. Two different integration methods were used for N-body simulations (with Mercury simulation package): Bulirsch-Stoer algorithm(bs) and hybrid methods. Each of these methods were run with three different timesteps:  $2.5\%$, $5.0\%$ and $10\%$ of the period of the innermost orbit.  In the right panel we show the relative difference in maximum eccentricity with respect to the maximum eccentricity computed by the different algorithms with the different timesteps. 

The hybrid integration scheme with different timesteps do not agree well when the eccentricity is high. Hence we chose to use the Burlisch-Stoer integration scheme for our N-body simulations. Burlisch-Stoer method agrees very well with the secular double averaged simulations for all the three timesteps that we chose, except when the configuration is near mean-motion resonances which are highlighted using dashed black lines. When alpha becomes larger than $\sim 0.18$, the system becomes chaotic (e.g., Figure \ref{fig:emaxecc}), so results at different timesteps with different integration methods no longer agree with each other. Numerical accuracy is high enough not to affect the maximum eccentricity.

\section{A2. Additional Surface of Sections}
\begin{figure*}
\includegraphics[width=\linewidth]{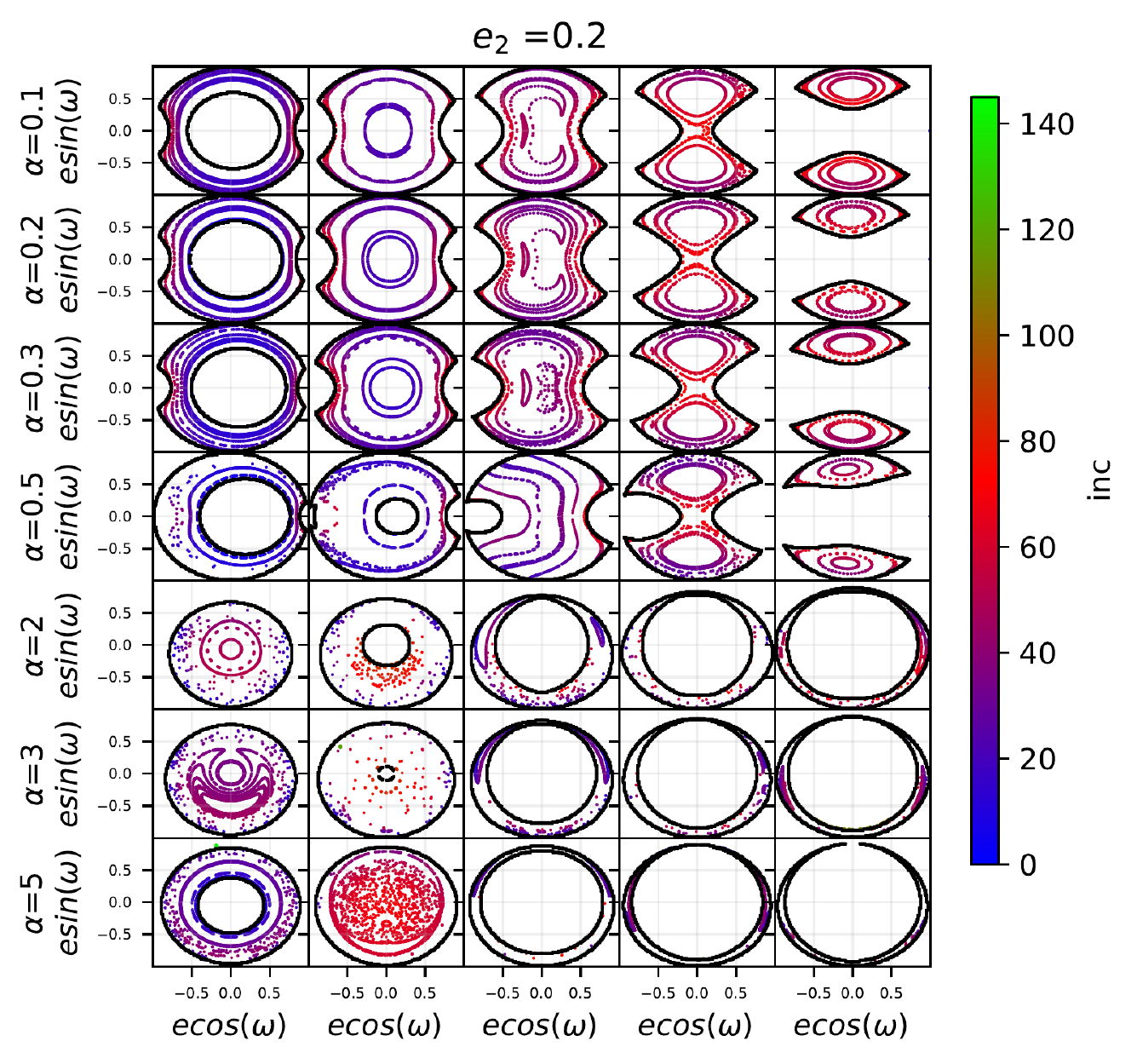}
	\caption{Surface of section in $e\cos(\omega)-e\sin(\omega)$ space for $e_2=0.2$. We use the conditions $\Omega=0$ and $\dot{\Omega}<0$ for inner test particle configurations and $\Omega=\pi/2$ and $\dot{\Omega}<0$ for outer test particle configurations to choose our points.}
\label{fig:e2}
\end{figure*}

\begin{figure*}[h]
\includegraphics[width=\linewidth]{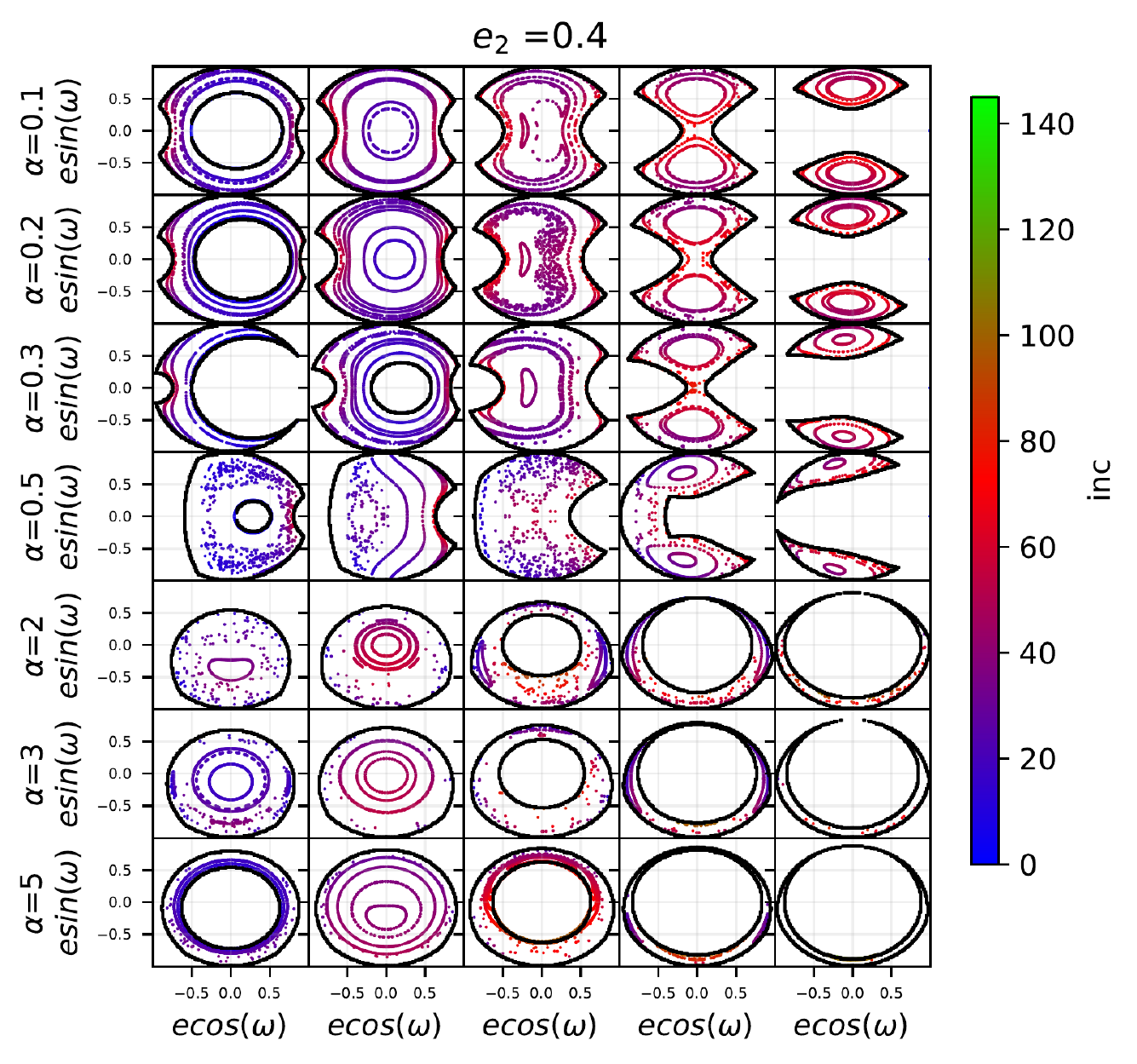}
	\caption{
	Same as Figure \ref{fig:i2}, except for $e_4=0.2$.  
	}
\label{fig:e4}
\end{figure*}

\begin{figure*}[h]
\includegraphics[width=\linewidth]{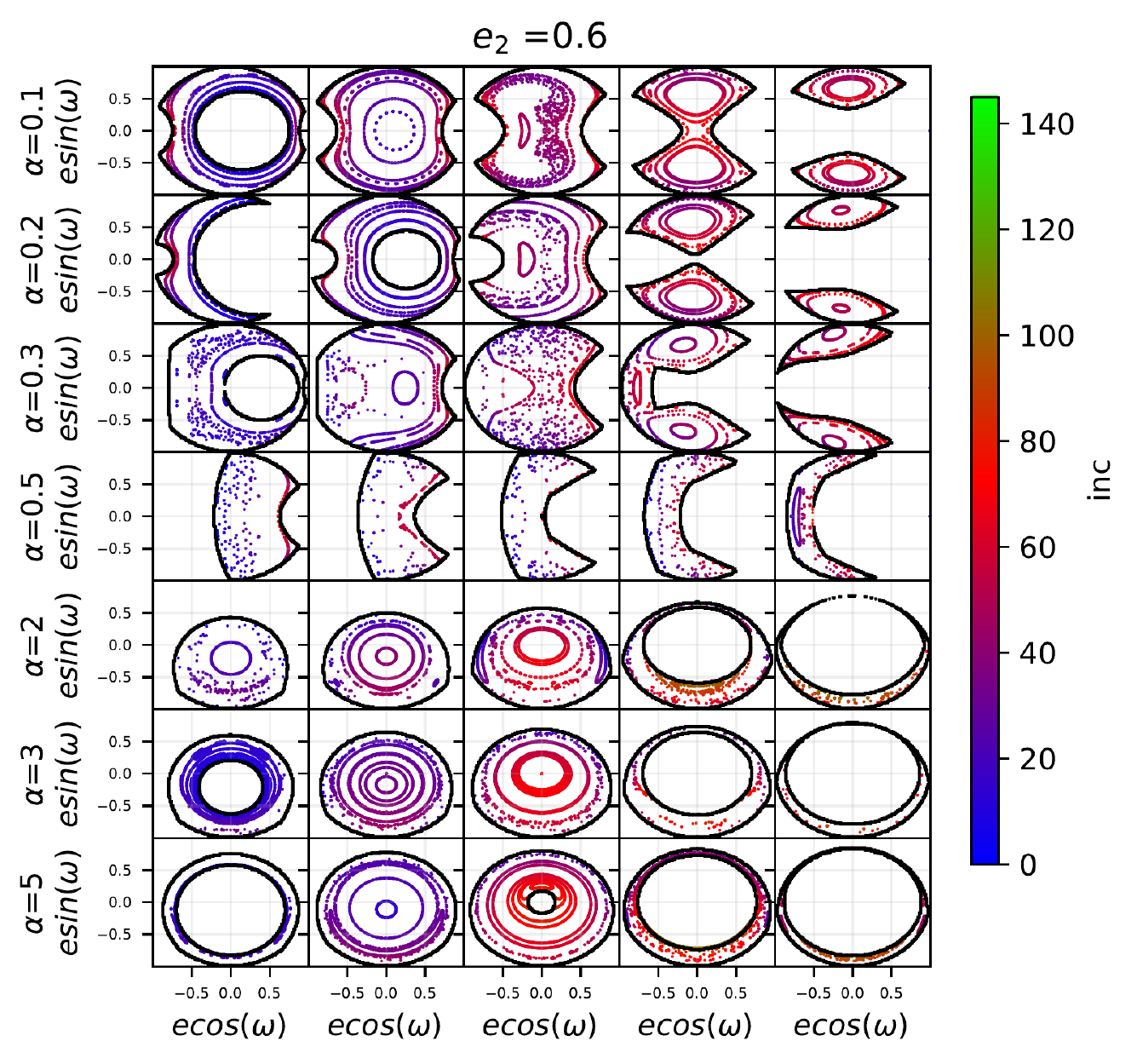}
	\caption{
	Same as Figure \ref{fig:i2}, except for $e_2=0.6$.  
	}
\label{fig:e6}
\end{figure*}

\begin{figure*}[h]
\includegraphics[width=\linewidth]{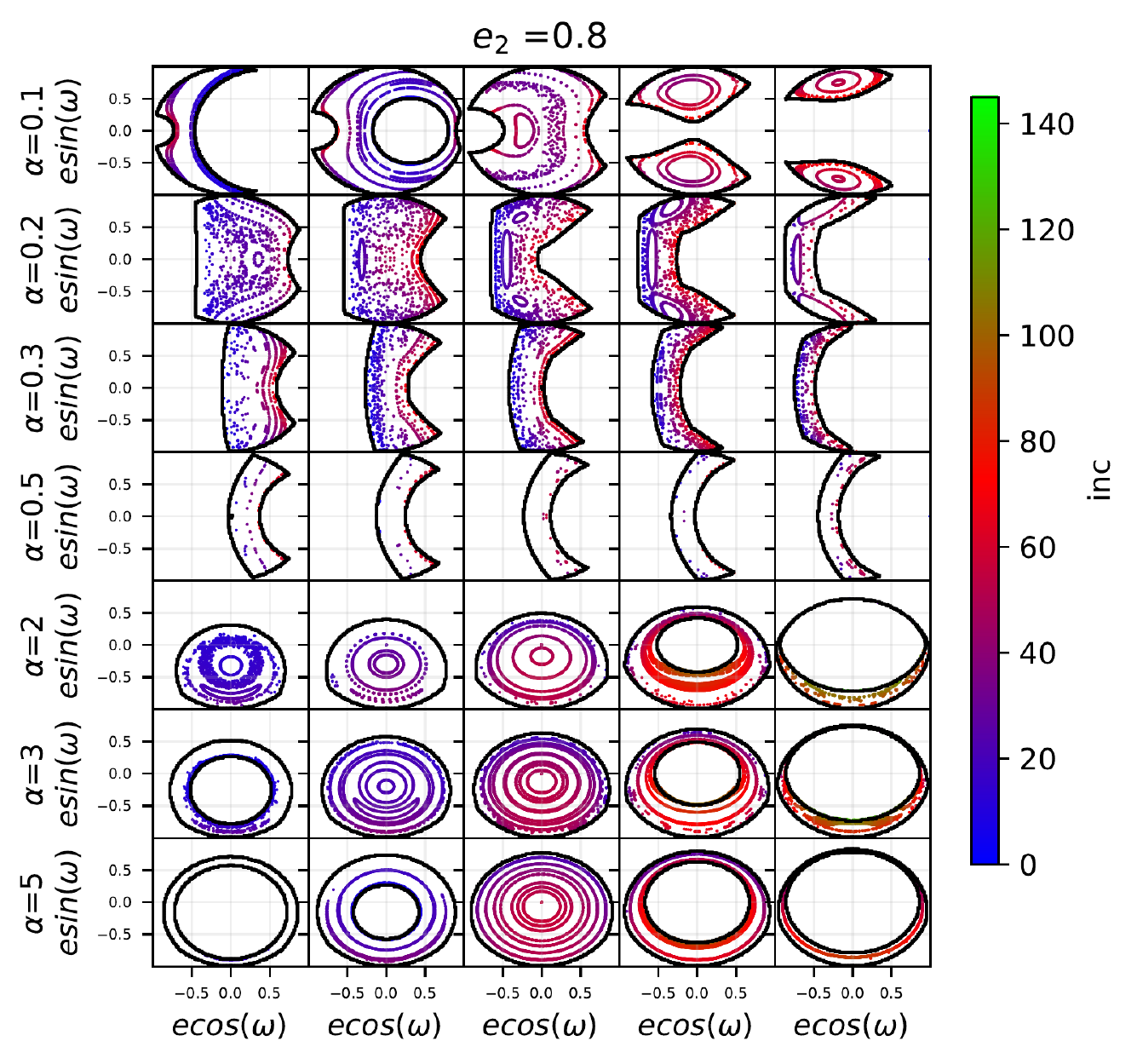}
	\caption{Same as Figure \ref{fig:i2}, except for $e_2=0.8$.  
	}
\label{fig:e8}
\end{figure*}

\begin{figure*}[h]
\includegraphics[width=\linewidth]{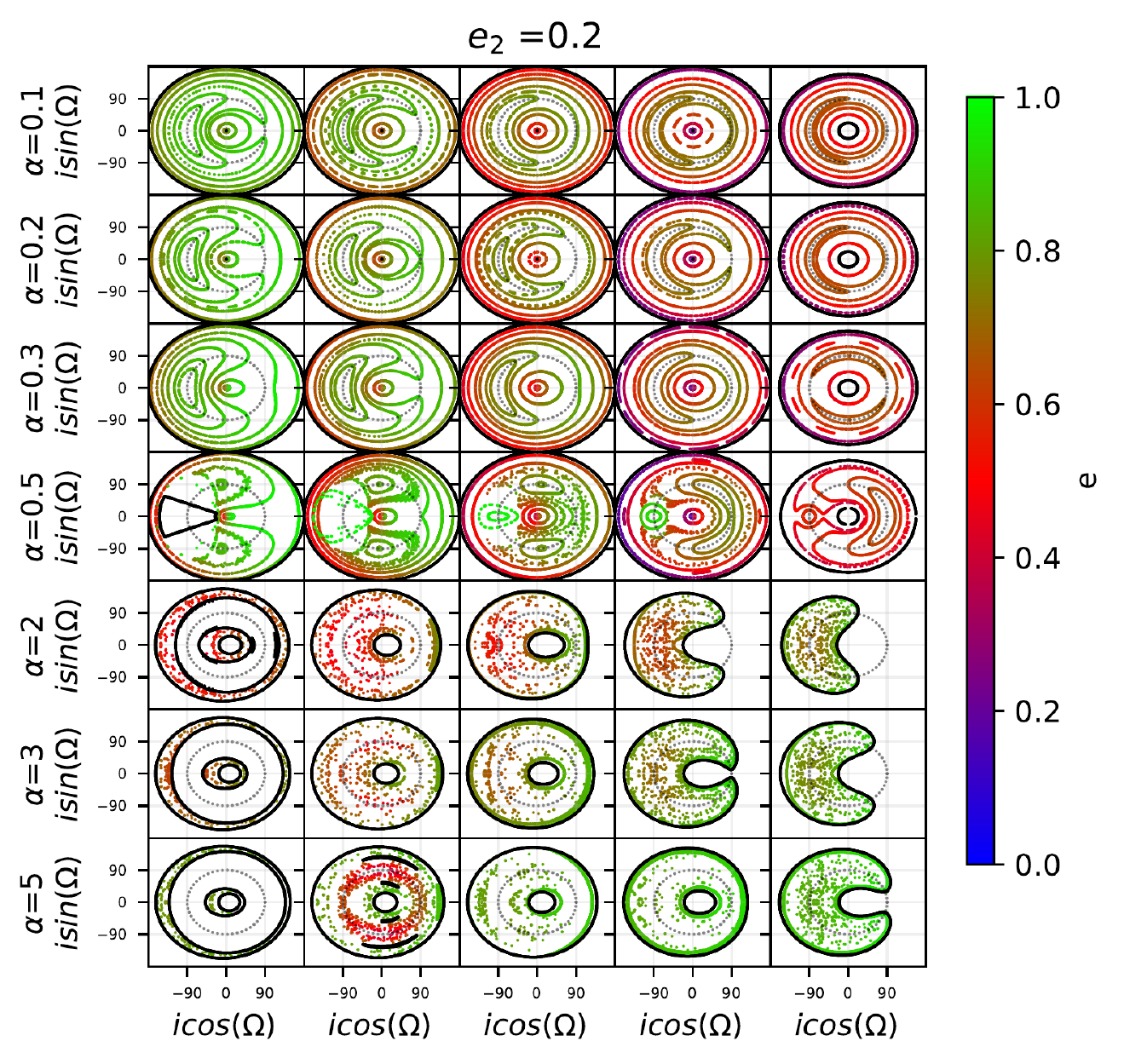}
	\caption{Surface of section in $i\cos(\Omega)-i\sin(\Omega)$ space for $e_2=0.2$. We use the conditions $\omega=0$ and $\dot{\omega}>0$ to choose our points. 
	}
\label{fig:i2}
\end{figure*}

\begin{figure*}[h]
\includegraphics[width=\linewidth]{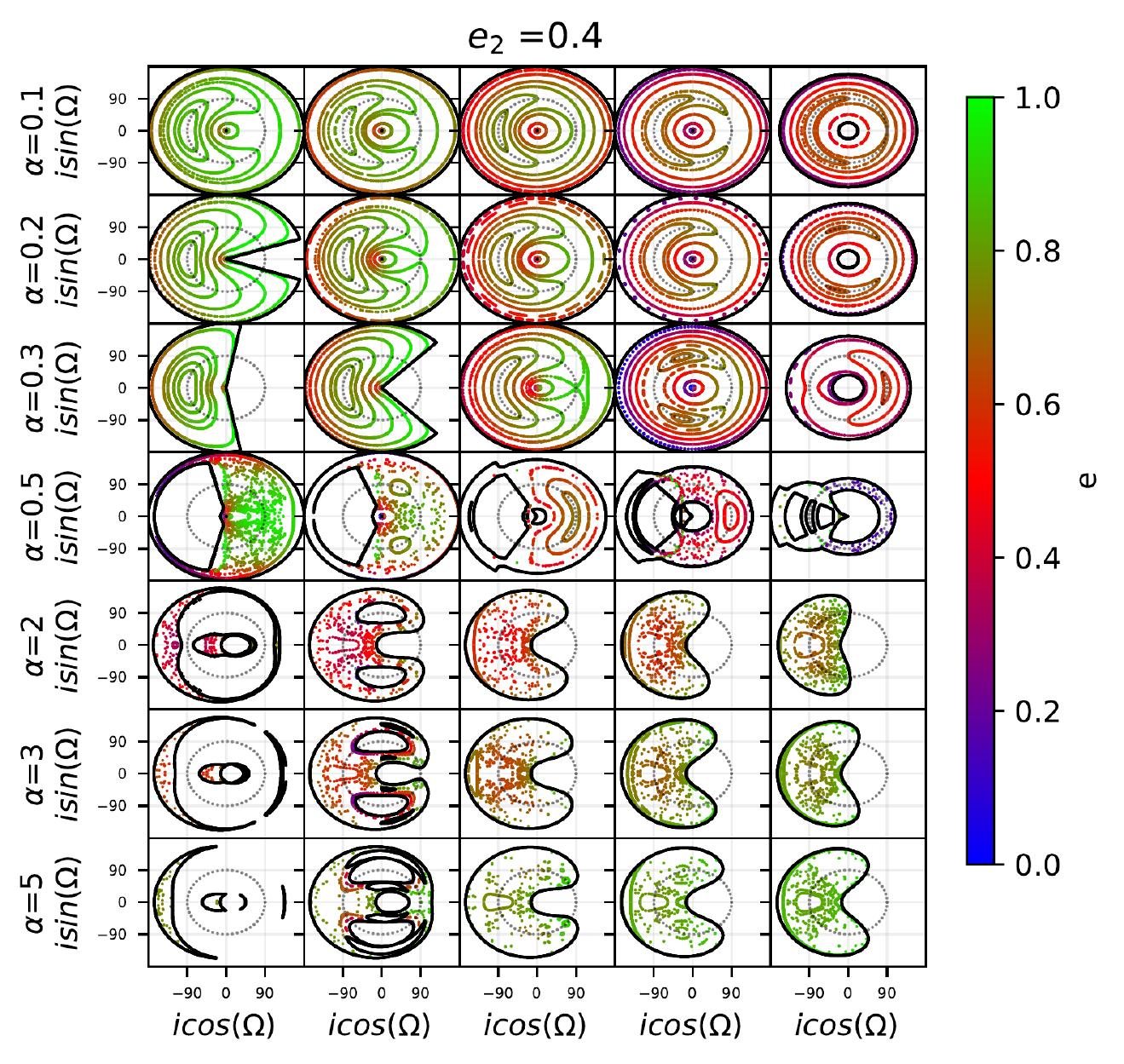}
	\caption{
	Same as Figure \ref{fig:e2}, except for $e_2=0.4$. 
	}
\label{fig:i4}
\end{figure*}

\begin{figure*}[h]
\includegraphics[width=\linewidth]{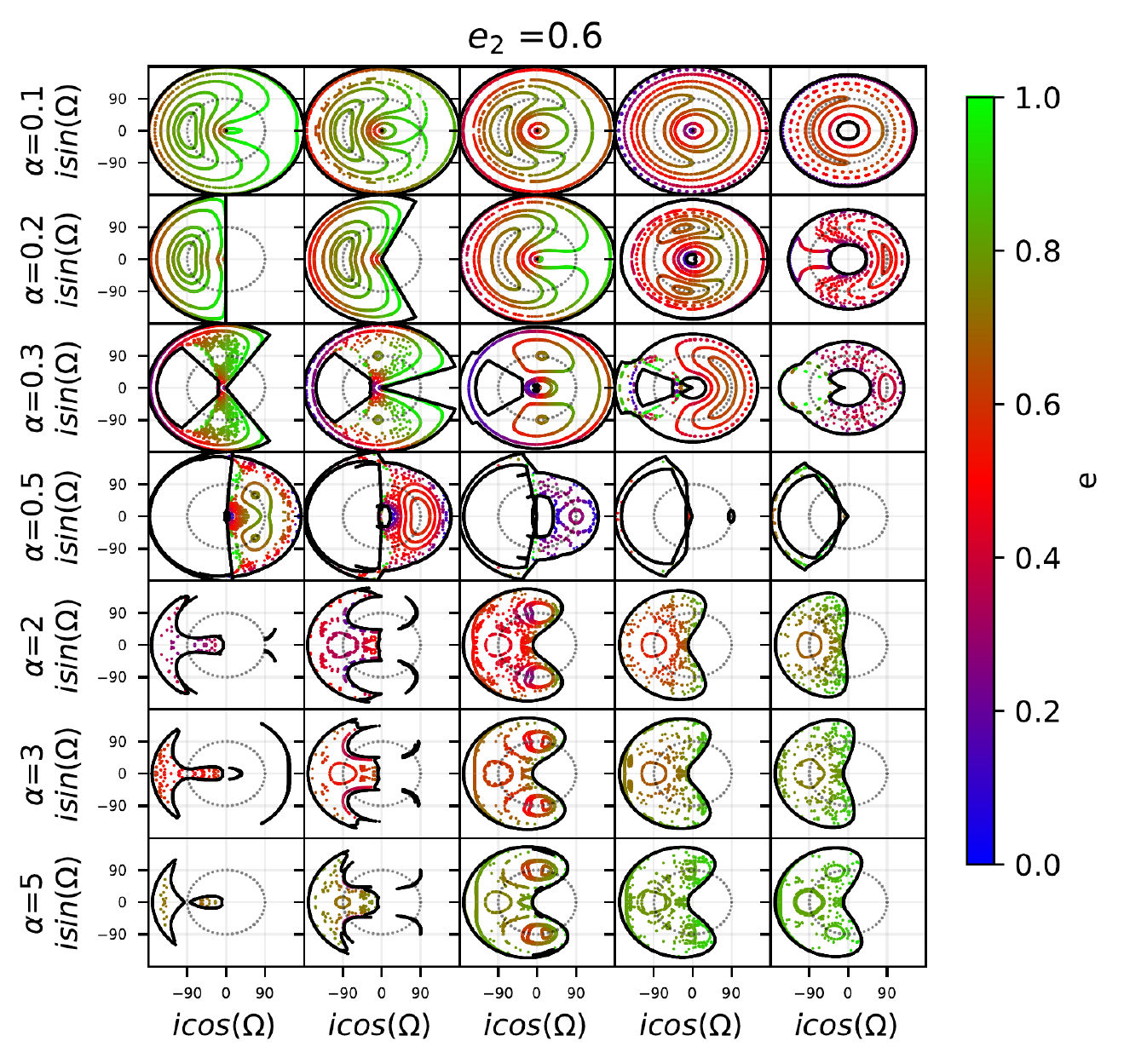}
\caption{
Same as Figure \ref{fig:e2}, except for $e_2=0.6$. 
}
\label{fig:i6}
\end{figure*}

\begin{figure*}[h]
\includegraphics[width=\linewidth]{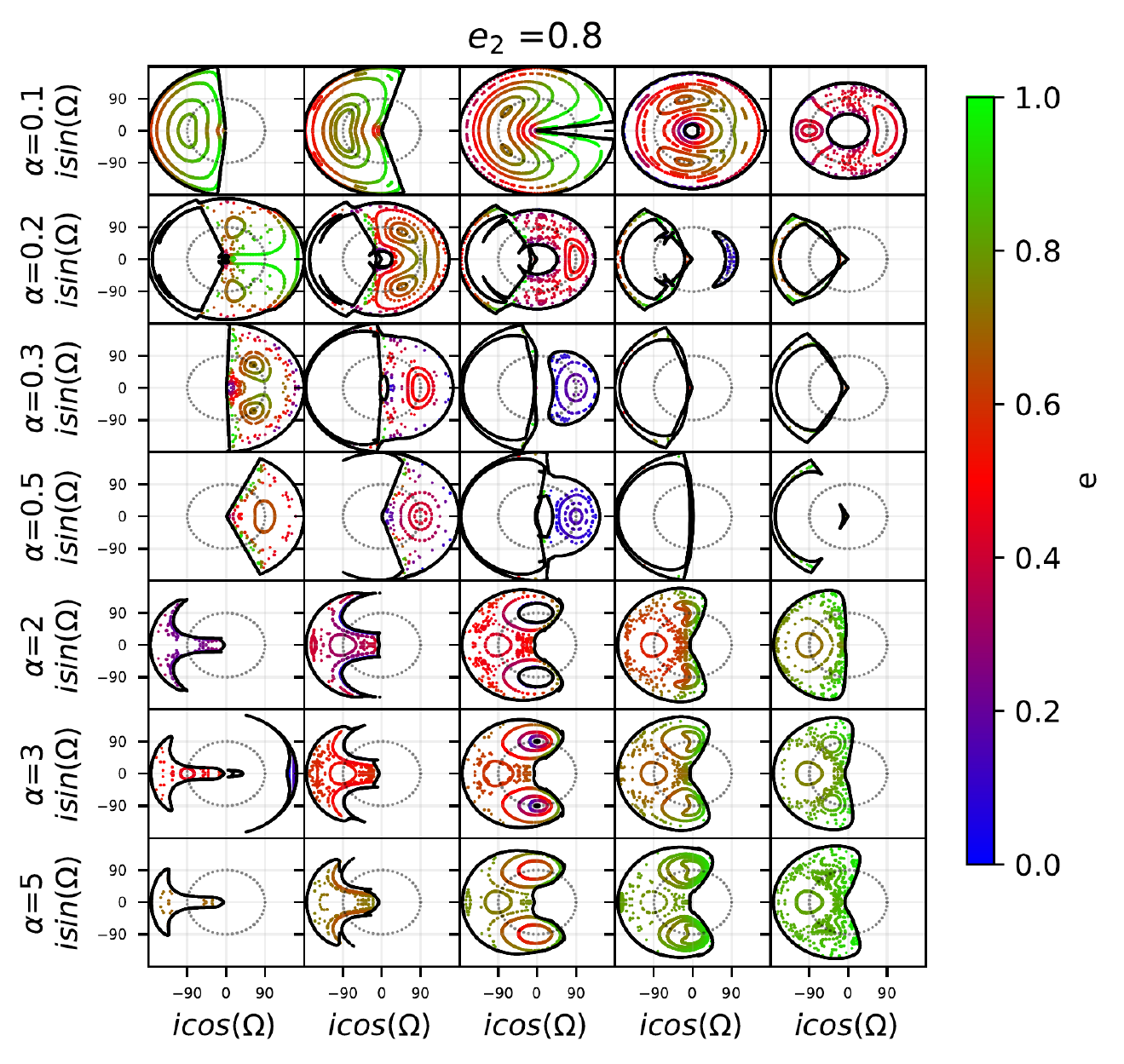}
\caption{Same as Figure \ref{fig:e2}, except for $e_2=0.8$. 
}

\label{fig:i8}
\end{figure*}

In this section we present all the surface of sections for different semi-major axis ratios $\alpha=\{0.1,0.2,0.3,0.5,2,3,5\}$ and perturbers eccentricities $e_2=\{0.2,0.4,0.6,0.8\}$. In Figures \ref{fig:e2}-\ref{fig:i8}, each row corresponds to a different $\alpha$, and each figure corresponds to a different outer binary eccentricity $e_2$ .  

For a given configuration of the system, only a finite range of Hamiltonian is physically allowed. To select the energy levels used to make the surfaces in $e-\omega$ space, we sample the values of $H$ on a grid in $e\cos(\omega)-e\sin(\omega)$ with $\Omega=0$. Then, we collect the $H$ values at the grid points, and choose the mean ($\langle H_{grids} \rangle$) as well as one and two standard deviations away from the mean ($\langle H_{grids} \rangle \pm \sigma_{Hgrids}$ and $\langle H_{grids} \rangle \pm 2\sigma_{Hgrids}$) to be the energy levels in the surfaces. The different columns in Figure \ref{fig:e2}-\ref{fig:i8} correspond to the different energy levels. We follow a similar procedure to select energy levels for surfaces in $i-\Omega$ space. 

By looking at different surface of sections in a given row (with the same semi-major axes ratio), it is clear that the dynamics depends sensitively on the value of the Hamiltonian. Eccentricities are generally excited to higher values when the energy is low for inner test particles, while it's the opposite when the test particle is on the outer orbit. Various features of these surfaces are presented in detail in Section \ref{section:soc}.

\bibliography{ref.bib}{}

\begin{thebibliography}{}
\expandafter\ifx\csname natexlab\endcsname\relax\def\natexlab#1{#1}\fi
\providecommand{\url}[1]{\href{#1}{#1}}

\bibitem[{Antonini {et~al.}(2014)Antonini, Murray, \&
  Mikkola}]{antonini_black_2014}
Antonini, F., Murray, N., \& Mikkola, S. 2014, {\textbackslash}apj, 781, 45

\bibitem[{Antonini \& Perets(2012)}]{antonini_secular_2012}
Antonini, F., \& Perets, H.~B. 2012, {\textbackslash}apj, 757, 27

\bibitem[{Bannister {et~al.}(2016)Bannister, Alexandersen, Benecchi, Chen,
  Delsanti, Fraser, Gladman, Granvik, Grundy, Guilbert-Lepoutre, Gwyn, Ip,
  Jakubik, Jones, Kaib, Kavelaars, Lacerda, Lawler, Lehner, Lin, Lykawka,
  Marsset, Murray-Clay, Noll, Parker, Petit, Pike, Rousselot, Schwamb,
  Shankman, Veres, Vernazza, Volk, Wang, \& Weryk}]{bannister_ossos._2016}
Bannister, M.~T., Alexandersen, M., Benecchi, S.~D., {et~al.} 2016,
  {\textbackslash}aj, 152, 212

\bibitem[{Bannister {et~al.}(2018)Bannister, Gladman, Kavelaars, Petit, Volk,
  Chen, Alexand~ersen, Gwyn, Schwamb, Ashton, Benecchi, Cabral, Dawson,
  Delsanti, Fraser, Granvik, Greenstreet, Guilbert-Lepoutre, Ip, Jakubik,
  Jones, Kaib, Lacerda, Van~Laerhoven, Lawler, Lehner, Lin, Lykawka, Marsset,
  Murray-Clay, Pike, Rousselot, Shankman, Thirouin, Vernazza, \&
  Wang}]{bannister_ossos._2018}
Bannister, M.~T., Gladman, B.~J., Kavelaars, J.~J., {et~al.} 2018,
  {\textbackslash}apjs, 236, 18

\bibitem[{Batygin {et~al.}(2019)Batygin, Adams, Brown, \&
  Becker}]{batygin_planet_2019}
Batygin, K., Adams, F.~C., Brown, M.~E., \& Becker, J.~C. 2019,
  {\textbackslash}physrep, 805, 1

\bibitem[{Batygin \& Brown(2016)}]{batygin_evidence_2016}
Batygin, K., \& Brown, M.~E. 2016, {\textbackslash}aj, 151, 22

\bibitem[{Beust(2016{\natexlab{a}})}]{beust_orbital_2016}
Beust, H. 2016{\natexlab{a}}, {\textbackslash}aap, 590, L2

\bibitem[{Beust(2016{\natexlab{b}})}]{beust_orbital_2016-1}
---. 2016{\natexlab{b}}, {\textbackslash}aap, 590, L2

\bibitem[{Boué \& Fabrycky(2014)}]{boue_compact_2014}
Boué, G., \& Fabrycky, D.~C. 2014, {\textbackslash}apj, 789, 110

\bibitem[{Chambers \& Migliorini(1997)}]{chambers_mercury_1997}
Chambers, J.~E., \& Migliorini, F. 1997, in {AAS}/{Division} for {Planetary}
  {Sciences} {Meeting} {Abstracts} \#29, {AAS}/{Division} for {Planetary}
  {Sciences} {Meeting} {Abstracts}, 27.06

\bibitem[{Cuk \& Burns(2004)}]{cuk_secular_2004}
Cuk, M., \& Burns, J.~A. 2004, {\textbackslash}aj, 128, 2518

\bibitem[{de~Elía {et~al.}(2019)de~Elía, Zanardi, Dugaro, \&
  Naoz}]{de_elia_inverse_2019}
de~Elía, G.~C., Zanardi, M., Dugaro, A., \& Naoz, S. 2019,
  {\textbackslash}aap, 627, A17

\bibitem[{{de La Fuente Marcos} \& {de La Fuente
  Marcos}(2014)}]{deLaFuenteMarcos14}
{de La Fuente Marcos}, C., \& {de La Fuente Marcos}, R. 2014, \mnras, 443, L59

\bibitem[{Fouvry {et~al.}(2017)Fouvry, Pichon, \&
  Magorrian}]{fouvry_secular_2017}
Fouvry, J.~B., Pichon, C., \& Magorrian, J. 2017, {\textbackslash}aap, 598,
  A71, \_eprint: 1606.05501

\bibitem[{{Gomes} {et~al.}(2015){Gomes}, {Soares}, \& {Brasser}}]{Gomes15}
{Gomes}, R.~S., {Soares}, J.~S., \& {Brasser}, R. 2015, \icarus, 258, 37

\bibitem[{{Grishin} {et~al.}(2018){Grishin}, {Perets}, \&
  {Fragione}}]{Grishin18}
{Grishin}, E., {Perets}, H.~B., \& {Fragione}, G. 2018, \mnras, 481, 4907

\bibitem[{Gronchi(2002)}]{gronchi_generalized_2002}
Gronchi, G.~F. 2002, Celestial Mechanics and Dynamical Astronomy, 83, 97.
\newblock \url{https://doi.org/10.1023/A:1020178613365}

\bibitem[{Gronchi \& Milani(1998)}]{gronchi_averaging_1998}
Gronchi, G.~F., \& Milani, A. 1998, Celestial Mechanics and Dynamical
  Astronomy, 71, 109

\bibitem[{{Gronchi} \& {Milani}(2001)}]{Gronchi01}
{Gronchi}, G.~F., \& {Milani}, A. 2001, \icarus, 152, 58

\bibitem[{Hadden {et~al.}(2018)Hadden, Li, Payne, \&
  Holman}]{hadden_chaotic_2018}
Hadden, S., Li, G., Payne, M.~J., \& Holman, M.~J. 2018, {\textbackslash}aj,
  155, 249

\bibitem[{Hamers(2018)}]{hamers_secular_2018}
Hamers, A.~S. 2018, {\textbackslash}mnras, 476, 4139, \_eprint: 1802.05716

\bibitem[{Hamers(2020)}]{hamers_secular_2020}
---. 2020, {\textbackslash}mnras, 494, 5492, \_eprint: 2004.08327

\bibitem[{Hamers \& Portegies~Zwart(2016)}]{hamers_secular_2016}
Hamers, A.~S., \& Portegies~Zwart, S.~F. 2016, {\textbackslash}mnras, 459,
  2827, \_eprint: 1511.00944

\bibitem[{Harrington(1968)}]{harrington_dynamical_1968}
Harrington, R.~S. 1968, {\textbackslash}aj, 73, 190

\bibitem[{{Henon}(1982)}]{Henon82}
{Henon}, M. 1982, Physica D Nonlinear Phenomena, 5, 412

\bibitem[{Katz {et~al.}(2011)Katz, Dong, \& Malhotra}]{katz_long-term_2011}
Katz, B., Dong, S., \& Malhotra, R. 2011, {\textbackslash}prl, 107, 181101

\bibitem[{Kozai(1962)}]{kozai_secular_1962}
Kozai, Y. 1962, {\textbackslash}aj, 67, 591

\bibitem[{{Li} \& {Adams}(2016)}]{Li_P9origin_16}
{Li}, G., \& {Adams}, F.~C. 2016, \apjl, 823, L3

\bibitem[{Li {et~al.}(2018)Li, Hadden, Payne, \& Holman}]{li_secular_2018}
Li, G., Hadden, S., Payne, M., \& Holman, M.~J. 2018, {\textbackslash}aj, 156,
  263

\bibitem[{Li {et~al.}(2014{\natexlab{a}})Li, Naoz, Holman, \&
  Loeb}]{li_chaos_2014}
Li, G., Naoz, S., Holman, M., \& Loeb, A. 2014{\natexlab{a}},
  {\textbackslash}apj, 791, 86

\bibitem[{Li {et~al.}(2014{\natexlab{b}})Li, Naoz, Kocsis, \&
  Loeb}]{li_eccentricity_2014}
Li, G., Naoz, S., Kocsis, B., \& Loeb, A. 2014{\natexlab{b}},
  {\textbackslash}apj, 785, 116

\bibitem[{Lidov(1962)}]{lidov_evolution_1962}
Lidov, M.~L. 1962, {\textbackslash}planss, 9, 719

\bibitem[{Lithwick \& Naoz(2011)}]{lithwick_eccentric_2011}
Lithwick, Y., \& Naoz, S. 2011, {\textbackslash}apj, 742, 94

\bibitem[{Luo {et~al.}(2016)Luo, Katz, \& Dong}]{luo_double-averaging_2016}
Luo, L., Katz, B., \& Dong, S. 2016, {\textbackslash}mnras, 458, 3060

\bibitem[{Millholland \& Laughlin(2019)}]{millholland_obliquity-driven_2019}
Millholland, S., \& Laughlin, G. 2019, Nature Astronomy, 3, 424

\bibitem[{Murray \& Dermott(1999)}]{murray_solar_1999}
Murray, C.~D., \& Dermott, S.~F. 1999, Solar system dynamics

\bibitem[{Murray \& Dermott(2000)}]{murray_solar_2000}
---. 2000, Solar {System} {Dynamics}

\bibitem[{Naoz(2016)}]{naoz_eccentric_2016}
Naoz, S. 2016, {\textbackslash}araa, 54, 441

\bibitem[{Naoz {et~al.}(2011{\natexlab{a}})Naoz, Farr, Lithwick, Rasio, \&
  Teyssandier}]{naoz_hot_2011-1}
Naoz, S., Farr, W.~M., Lithwick, Y., Rasio, F.~A., \& Teyssandier, J.
  2011{\natexlab{a}}, {\textbackslash}nat, 473, 187

\bibitem[{Naoz {et~al.}(2011{\natexlab{b}})Naoz, Farr, Lithwick, Rasio, \&
  Teyssandier}]{naoz_hot_2011}
---. 2011{\natexlab{b}}, {\textbackslash}nat, 473, 187

\bibitem[{Naoz {et~al.}(2013)Naoz, Farr, Lithwick, Rasio, \&
  Teyssandier}]{naoz_secular_2013}
---. 2013, {\textbackslash}mnras, 431, 2155

\bibitem[{Naoz {et~al.}(2017)Naoz, Li, Zanardi, de~Elía, \&
  Di~Sisto}]{naoz_eccentric_2017}
Naoz, S., Li, G., Zanardi, M., de~Elía, G.~C., \& Di~Sisto, R.~P. 2017,
  {\textbackslash}aj, 154, 18

\bibitem[{{Nesvold} {et~al.}(2016){Nesvold}, {Naoz}, {Vican}, \&
  {Farr}}]{nesvold_2016}
{Nesvold}, E.~R., {Naoz}, S., {Vican}, L., \& {Farr}, W.~M. 2016, \apj, 826, 19

\bibitem[{{Saillenfest} {et~al.}(2017){Saillenfest}, {Fouchard}, {Tommei}, \&
  {Valsecchi}}]{saillenfest17}
{Saillenfest}, M., {Fouchard}, M., {Tommei}, G., \& {Valsecchi}, G.~B. 2017,
  Celestial Mechanics and Dynamical Astronomy, 129, 329

\bibitem[{Saillenfest {et~al.}(2017)Saillenfest, Fouchard, Tommei, \&
  Valsecchi}]{saillenfest_non-resonant_2017}
Saillenfest, M., Fouchard, M., Tommei, G., \& Valsecchi, G.~B. 2017, Celestial
  Mechanics and Dynamical Astronomy, 129, 329

\bibitem[{Saillenfest \& Lari(2017)}]{saillenfest_long-term_2017}
Saillenfest, M., \& Lari, G. 2017, {\textbackslash}aap, 603, A79

\bibitem[{Shankman {et~al.}(2017)Shankman, Kavelaars, Lawler, Gladman, \&
  Bannister}]{shankman_consequences_2017}
Shankman, C., Kavelaars, J.~J., Lawler, S.~M., Gladman, B.~J., \& Bannister,
  M.~T. 2017, {\textbackslash}aj, 153, 63

\bibitem[{Sheppard \& Trujillo(2016)}]{sheppard_new_2016}
Sheppard, S.~S., \& Trujillo, C. 2016, {\textbackslash}aj, 152, 221

\bibitem[{{Stone} \& {Leigh}(2019)}]{Stone19}
{Stone}, N.~C., \& {Leigh}, N. W.~C. 2019, \nat, 576, 406

\bibitem[{Touma {et~al.}(2009)Touma, Tremaine, \&
  Kazandjian}]{touma_gausss_2009}
Touma, J.~R., Tremaine, S., \& Kazandjian, M.~V. 2009, {\textbackslash}mnras,
  394, 1085

\bibitem[{Valtonen \& Karttunen(2006)}]{valtonen_three-body_2006}
Valtonen, M., \& Karttunen, H. 2006, The {Three}-{Body} {Problem}

\bibitem[{Vinson \& Chiang(2018)}]{vinson_secular_2018}
Vinson, B.~R., \& Chiang, E. 2018, {\textbackslash}mnras, 474, 4855

\bibitem[{Yokoyama {et~al.}(2003)Yokoyama, Santos, Cardin, \&
  Winter}]{yokoyama_orbits_2003}
Yokoyama, T., Santos, M.~T., Cardin, G., \& Winter, O.~C. 2003,
  {\textbackslash}aap, 401, 763

\bibitem[{{Zanardi} {et~al.}(2017){Zanardi}, {de El{\'\i}a}, {Di Sisto},
  {Naoz}, {Li}, {Guilera}, \& {Brunini}}]{Zanardi17}
{Zanardi}, M., {de El{\'\i}a}, G.~C., {Di Sisto}, R.~P., {et~al.} 2017, \aap,
  605, A64

\end{thebibliography}
\bibliographystyle{aasjournal}

\end{document}